\documentstyle[psfig]{l-aa}

\def\mic {$\mu\hbox{m}$}
\def\um {$\mu\hbox{m}$}

\def\kms {\hbox{${\rm km\ts s}^{-1}$}}
\def\percc {$\hbox{{\rm cm}}^{-3}$}    
\def\cmsq  {$\hbox{{\rm cm}}^{-2}$}    
\def\arcsec {\hbox{$^{\prime\prime}$}}
\def\arcmin {\hbox{$^{\prime}$}}

\def\Av {A$_V$}
\def\MOLH {\hbox{${\rm H}_2$}}  
\def\CEIO {\hbox{${\rm C}^{18}{\rm O}$}}   
\def\Lsun {$\hbox{L}_\odot$}
\def\Tstar {T$_\star$}
\def\Lstar {L$_\star$}
\def\cmsq  {$\hbox{{\rm cm}}^{-2}$}    
\def\THEC  {$\Theta ^{1}$C Ori}  
\def\THEA  {$\Theta ^{2}$A Ori}  
\def\sbu {erg cm$^{-2}$ s$^{-1}$ sr$^{-1}$}
\def\simless{\mathbin{\lower 3pt\hbox
      {$\rlap{\raise 5pt\hbox{$\char'074$}}\mathchar"7218$}}} 
\def\simgreat{\mathbin{\lower 3pt\hbox
     {$\rlap{\raise 5pt\hbox{$\char'076$}}\mathchar"7218$}}} 

\def\OIa {2p$^3$3D-2p$^3$3P}
\def\OIb {2p$^3$4S-2p$^3$3P}
\begin{document}
\thesaurus{08(09.03.1; 09.13.2; 13.19.3)}

\title{Near Infrared Spectra of the
Orion Bar}

\author{A.~Marconi \inst{1,3} \and L.~Testi \inst{1,4} 
\and A.~Natta \inst{2} \and  C.M.~Walmsley  \inst{2} }

\institute{
    Dipartimento di Astronomia e Scienza dello Spazio, Universit\`a
    degli Studi di Firenze, Largo E.Fermi 5, I-50125 Firenze, Italy
   \and
    Osservatorio Astrofisico di Arcetri, Largo E.Fermi 5,
    I-50125 Firenze, Italy
   \and
    Space Telescope Science Institute, 3700 San Martin Drive, 
    Baltimore, MD 21218
    \and
    Division of Physics Mathematics and Astronomy, Caltech,
    MS 105-24, Pasadena, CA 91125, USA
           }
         
\offprints{C.M. Walmsley}
\date{Received date; accepted date}
\maketitle
\markboth{A.~Marconi et al.:
          NIR Spectra of the Orion Bar}
%
%
\bigskip
%
%
\begin{abstract}
 We have used the LONGSP spectrometer on the 1.5-m TIRGO telescope to
 obtain  long slit spectra in the J, H, and K  wavelength bands
 towards two positions along the Orion bar. These data have
 been supplemented with images made using the ARNICA camera mounted
 on TIRGO as well as with an ESO NTT observation carried out by
 Dr A. Moorwood. We detect a variety of transitions of hydrogen,
 helium, OI, FeII, FeIII, and \MOLH .  From our molecular hydrogen
 data, we conclude that densities are moderate ($3-6\times  10^4$
 \percc ) in the layer responsible for the 
  molecular hydrogen emission and
 give no evidence for the presence of dense neutral clumps. We
 also find that the molecular hydrogen bar is likely to be
 tilted by $\sim$10 degrees relative to the line of sight. 
 We discuss the relative merits of several models of the structure of
 the bar and conclude that it may be split into two 
  structures separated by 0.2-0.3 parsec along the line of sight. 
  It also seems likely to us that in both structures,
  density increases along a line
  perpendicular to the ionization front which penetrates into
  the neutral gas.  
  
  We have used the 1.317\mic \ OI line to  estimate the FUV
  radiation field incident at the ionization front and find
  values of $1-3\times 10^4$ greater than
  the  average interstellar field.
 From  [FeII] line measurements, 
 we conclude that the electron density
 in the ionized layer associated with the ionization front is of
 order $10^4$ \percc.
Finally, our analysis of the helium and
 hydrogen recombination lines implies essential coincidence of the
 helium and hydrogen Str\"{o}mgren spheres. 
 \keywords{interstellar medium: HII regions --
 interstellar medium : Orion -- IR: interstellar medium : lines and bands}
\end{abstract}
%
%
\section{Introduction}


 The properties of the Orion nebula   are the starting point
 for many of our ideas on high mass stars and their interactions with
 the environment.  A symposium held in 1981 (Glassgold et al.~\cite{Gea82})
 summarises much of what was known at that time. More recent work has
 been reviewed by Genzel \& Stutzki~(\cite{GS89}).  In general, the aim
 has been to understand the ionization structure and dynamical evolution
 of the nebula. In recent years, much attention has been paid to 
 the hot neutral gas  adjacent to the ionization front known as
 a PDR or Photon Dominated Region.

  Work on the ionized nebula has tended to
  concentrate upon determinations of the ionization structure and
 elemental abundances  (see Peimbert~\cite{P82}, Simpson et al.~\cite{Sea86}). 
 More recent optical and infrared studies 
  have been carried out by Osterbrock et
 al.~(\cite{OSV90}, \cite{OTV92}), Baldwin et al.~(\cite{Bea91}),  Peimbert et al.~(\cite{PTR92}),
 Pogge et al.~(\cite{POA92}), De Poy \& Pogge~(\cite{DP94}),
 Bautista et al.~(\cite{BPD95}), 
 Rubin et al.~(\cite{RDW93}), and Rodriguez (1996). 
 A review of the results has been
 made by Peimbert~(\cite{P93}) and discussions of the methods
 employed are given by Mathis~(\cite{M95}) and by Peimbert~(\cite{P95}). These studies in
 general indicate that the major fraction  of elements such as
 C,N,O,S are in the gas phase within the ionized nebula 
 whereas species such as Si and Fe  appear to be depleted by
 roughly an order of magnitude relative to abundances either in the
 Sun or nearby B stars. 

 Radio work on Orion  has revealed an immense variety of
 structures in the emissions of the ionized gas ( e.g. Felli
 et al.~\cite{Fea93}, Yusuf--Zadeh~\cite{YZ90}).  Particularly striking is 
 the bar--like structure situated roughly 2 arc minutes
 (0.25 parsec) to the south--east of the Trapezium stars 
 which is the subject of this article. ``The Bar'' is also
 observed in molecular line emission  (see below) 
 and clearly marks an ionization front where Lyman continuum photons
 from the O6 star 
 \THEC  \
 are absorbed.  The dynamical behavior of the
 ionized gas can be studied in radio recombination lines
 (Pankonin et al.~\cite{PWH79}, Wilson \& J\"{a}ger~\cite{WJ87}, Wilson \&
 Filges~\cite{WF90}, Wilson et al. ~\cite{WFCRR97}) 
 from which one concludes that much of the ionized material
 in Orion is streaming towards the observer. 

  Infared studies of HII regions sample not only the ionized
  gas but also the adjacent neutral material or PDR. 
 A recent review  of the properties of these regions
 is that of Hollenbach \& Tielens~(\cite{HT97}) (see also the discussions of
 Genzel~\cite{G92},  and Walmsley~\cite{W97}). Modelling
 studies have been carried out by Tielens \& Hollenbach~(\cite{TH85}),
 Hollenbach et al.~(\cite{HTT91}), Sternberg \& Dalgarno~(\cite{SD89}; \cite{SD95}),
 Fuente et al.~(\cite{FUea93}),  Jansen et al.~(\cite{Jea95a},b),
 Bertoldi \& Draine ~(\cite{BD96}) and
 Draine \& Bertoldi ~(\cite{DB96}).  
 Much of this activity has centred on attempts to understand
 the properties of "The Bar" mentioned above. 
  Recent observational studies  using
 a variety of molecular tracers have been carried out
 by Tielens et al.~(\cite{Tea93}), by Tauber et al.~(\cite{Tea94}; \cite{tauber95}), by Hogerheijde
 et al.~(\cite{HJD95}), and by van der Werf et al.~(\cite{VdW96}).  These show a
 stratification  along the direction of the perpendicular to the
 bar in the plane of the sky.  This is in the sense expected for
 gas heated by the Trapezium stars and consistent according to the
 models with attenuation by a gas of density $5\, 10^4$ \percc .
 However, the data also seem to
 show that the gas in the bar is far from homogeneous and that
 clumps of density as high as $10^6$ \percc  \ are embedded in the
 filament. Such high density condensations presumably either have been
 or will be soon overrun by the ionization front and will give rise
 to dense ionized globules within the HII region (see e.g Lizano et al.
 ~\cite{LCGH96}, Dyson et al. ~\cite{DWR95}). 
 Understanding the characteristics of such high density
 clumps may thus be of critical importance for the evolution of the
 HII region. 

   One of the most useful tracers of PDR's has turned out to be the
   near infrared lines of molecular hydrogen. 
   For example, van der Werf et al. used the FAST 
 camera on the ESO/MPI 2.2 m telescope 
 to image the \MOLH \ $v=1\rightarrow 0$ S(1)  (2.122 \mic )
  and $v=2\rightarrow 1$ S(1) (2.248 \mic ) lines towards the
  bar with 1.5\arcsec \ resolution.  These show that the transition
  from atomic to molecular hydrogen in the bar occurs 15 arc seconds 
  (0.03 pc at a distance of 450 pc) to the SE of the ionization front 
  (i.e. away from the ionizing stars).  Van der Werf et al. also find
  that the ratio $R_{12}$ of the intensities  of the $1\rightarrow 0$
  and $2\rightarrow 1$ lines varies between a value of $8.1\pm 0.7$
  at the peak of the \MOLH \ emission to a value of $3.4\pm 1.9$
  30\arcsec \ from the ionization front on the side shielded from
  the radiation of the Trapezium stars. The latter value is
  characteristic of UV-pumped fluorescent emission in a low density
  gas (Sternberg and Dalgarno~\cite{SD89}). 

\begin{figure*}
\centerline{\psfig{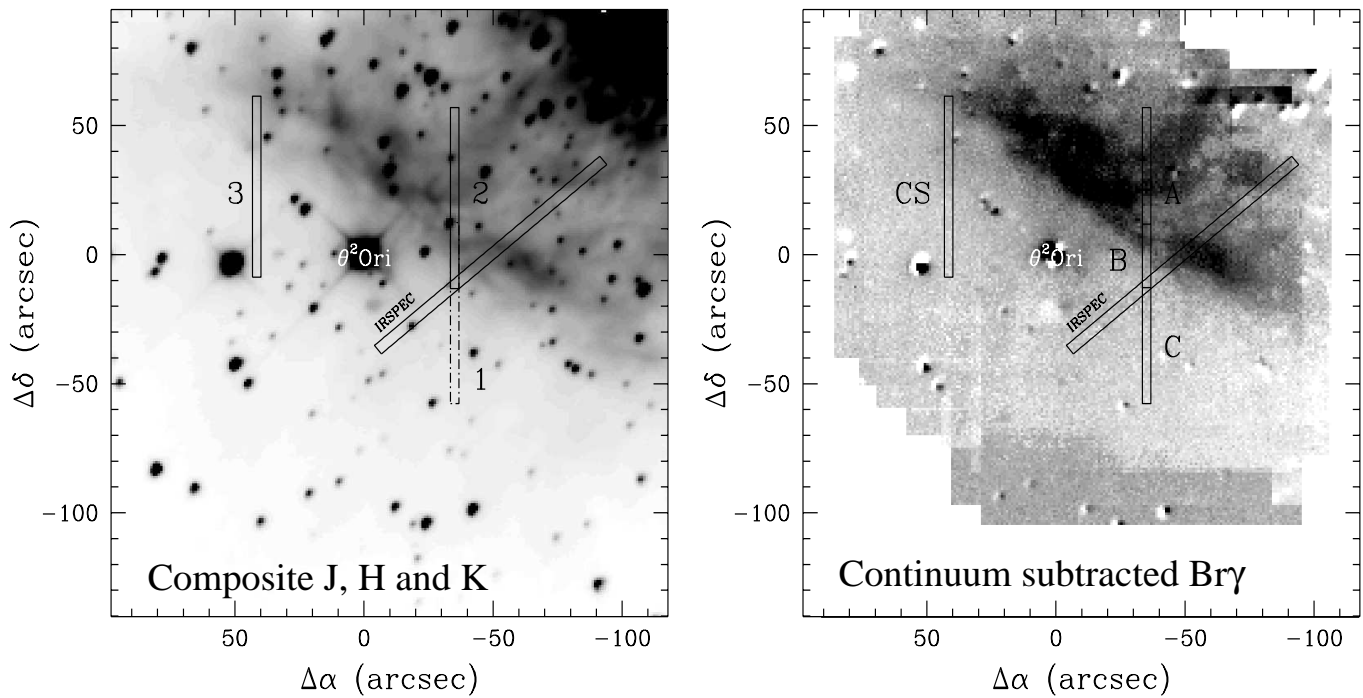}}
\caption
  {On the left: composite J, H, and K image of the Orion Bar region showing the slit positions used for the LONGSP and IRSPEC observations. Coordinates are
  offsets in right ascension and declination
  relative to the position of the star \THEA \ 
  (R.A(1950)= 5$^{h}$ 32$^{m}$ 55.$^{s}$5 , Dec(1950)=
  -5$^{\circ}$ 26\arcmin \ 51\arcsec ). We show our three slit
  positions (1,2,and 3) as well as the four positions for which
  we tabulate line intensities (A,B,C,CS). On the right:
  Br$\gamma$\
  image of the Orion Bar region.}
\label{fima}
\end{figure*}

  The present study had as its aim to obtain near IR spectra of the
  gas in the vicinity of the bar in order to verify and extend
 understanding of the physical conditions on both sides of the 
 ionization front. 
 We were partly motivated by the idea that there is a link  between the
 ionized and PDR components in that the former is mainly sensitive to the
 radiation just shortward of 912 \AA \ while the latter is basically a
 measure of the radiation longward of this limit (see  Bertoldi \& Draine
 1996 for a discussion).  It is thus of considerable interest to compare
 the two using the same instrument.  We therefore used the TIRGO
 telescope on the Gornergrat (Switzerland) to obtain slit spectra
 in the  J, H,  and K bands at 3 positions in the vicinity of the bar,
 shown in Fig.~\ref{fima}.
 As supplementary information, we also made use of
 unpublished observations carried out using the IRSPEC spectrometer on the
 ESO NTT telescope by Dr A.Moorwood. 
 
 We summarize in the next section the  techniques used for
 observations and data reductions. The results are
 presented in Sect.3 and discussed in Sect. 4.
 We summarize our
 conclusions in Sect. ~\ref{sconcl}.

\begin{table}
\begin{center}
\caption{Spectra in the J, H and K bands}
\vskip 0.2cm
\vbox{\hskip -8mm
\begin{tabular}{lccccc}
\hline\hline
Line & $\lambda$  & $A$ & $B$ &  $C$ & $CS$ \\
     & ($\mu$m)   &  &  &  &  \\
\hline
&&&&&\\
\,[OI]\,$2p^33D-2p^33P$          & 1.129 & 3.6 & 4.7 & 6.8 & 12 \\
\,[PII]\,$3p^2D_2-3p^2P_2$        & 1.189 & 2.4 & 3.5 & 3.2 & 6.6 \\
\,HeI\,$5^3D-3^3P^\circ$	 & 1.197 & 5.1 & 5.4 & 6.0 & 5.4 \\
\,HeI\,$4^3P^\circ-3^3S$     & 1.253 & 6.9 & 5.9 & 5.3 & 2.3 \\
\,[FeII]\,$4sa^4\!D_{7/2}\!-\!4sa^6\!D_{9/2}$& 1.257 & 4.8 & 8.3 & 6.7 & 18 \\
\,??                             & 1.268 & 2.3 & 1.5 & $<$1.8 & 5.7 \\
\,HeI\,$5^3F^\circ-3^3D$     & 1.278 & 18 & 16 & 18 & 26 \\
\,HI\,5-3                        & 1.282 & 370 & 400 & 450 & 430 \\
\,HeI\,$5^3P^\circ-3^3D$     & 1.298 & 2.2 & 2.3 & 1.6 & $<$2.6 \\
\,+[FeII]\,$4sa^4\!D_{3/2}\!-\!4sa^6\!D_{1/2}$&     &     &       &       &       \\
\,[OI]\,$2p^34S-2p^33P$          & 1.317 & 3.3 & 6.8 & 7.4 & 8.4 \\
\,[FeII]\,$4sa^4\!D_{7/2}\!-\!4sa^6\!D_{7/2}$ & 1.321 & 1.1 & 2.3 & $<$1.6 & 4.7 \\
\,HeI\,$5^1S-3^1P^\circ$     & 1.341 & 1.5 & 1.3 & 1.6 & 2.3 \\
&&&&&\\
\,HI\,21-4                      & 1.514 & 3.3 & 2.7 & 2.7 & 3.0 \\
\,HI\,20-4                      & 1.520 & 3.6 & 3.6 & 3.6 & 3.9 \\
\,HI\,19-4                      & 1.526 & 4.4 & 4.3 & 4.5 & 4.4 \\
\,HI\,18-4                      & 1.534 & 6.4 & 6.2 & 6.6 & 8.7 \\
\,+[FeII]\,$4sa^4\!D_{5/2}\!-\!3d^7\!a^4\!F_{9/2}$ &     &     &       &       &       \\
\,HI\,17-4                      & 1.544 & 6.5 & 5.9 & 6.3 & 6.2 \\
\,HI\,16-4                      & 1.556 & 8.2 & 7.3 & 7.2 & 7.2 \\
\,HI\,15-4                      & 1.570 & 9.4 & 9.0 & 8.2 & 9.4 \\
\,HI\,14-4                      & 1.588 & 11 & 10 & 10 & 13 \\
\,[FeII]\,$4sa^4\!D_{3/2}\!-\!3d^7\!a^4\!F_{7/2}$& 1.600 & 0.5 & 0.6 & 0.9 & 3.0 \\
\,HI\,13-4                      & 1.611 & 14 & 13 & 13 & 16 \\
\,HI\,12-4                      & 1.641 & 18 & 17 & 17 & 19 \\
\,[FeII]\,$4sa^4\!D_{7/2}\!-\!3d^7\!a^4\!F_{9/2}$& 1.644 & 5.9 & 11 & 8.7 & 24 \\
\,[FeII]\,$4sa^4\!D_{5/2}\!-\!3d^7\!a^4\!F_{7/2}$& 1.677 & 0.8 & 1.1 & $<$1.1 & 2.3 \\
\,HI\,11-4                      & 1.681 & 21 & 21 & 23 & 24 \\
\,HeI\,$4\,^3D-3\,^3P^\circ$    & 1.701 & 9.1 & 8.5 & 8.6 & 6.4 \\
\,HeI\,$10\,^1P^\circ-4\,^1D$   & 1.732 & 0.5 & 1.0 & 2.1 & 1.2 \\
\,HI\,10-4                      & 1.737 & 26 & 28 & 29 & 30 \\
\,HeI\,$7^3P^\circ-4^3S$    & 1.746 & 0.9 & 1.8 & 3.0 & 3.9 \\
$\>\>$ +H$_2$(1,0)S(7)          &     &     &       &       &       \\
\,H2(1,0)S(6)                   & 1.788 & 1.5 & 1.1 & 1.6 & 2.2 \\
&&&&&\\
\,H$_2$(1,0)S(2)                & 2.034 & 0.6 & .... & .... & 4.9 \\
\,HeI\,$2^1P^\circ-2^1S$    & 2.058 & 99 & 110 & 74 & 95 \\
\,H$_2$(2,1)S(3)                & 2.073 & $<$0.5 & 1.2 & 3.6 & 4.2 \\
\,HeI\,$4S-^3P^\circ$           & 2.113 & 4.2 & 3.7 & 2.8 & 2.1 \\
\,H$_2$(1,0)S(1)                & 2.122 & 1.5 & 5.1 & 17 & 15 \\
\,H$_2$(2,1)S(2)                & 2.154 & $<$0.1 & 0.2 & 1.3 & 1.3 \\
\,HeI\,$7F^\circ-4^1D$          & 2.162 & 3.0 & 2.5 & 2.4 & 1.9 \\
\,HI\,7-4                       & 2.166 & 100 & 100 & 100 & 100 \\
\,??+H$_2$(3,2)S(3)             & 2.199 & 0.5 & 0.9 & 1.4 & 3.5 \\
\,[FeIII]$3d^6G_5-3D^6H_6$      & 2.219 & 1.4 & 2.0 & 1.3 & 2.3 \\
\,H$_2$(1,0)S(0)                & 2.223 & 0.5 & 1.6 & 6.6 & 6.5 \\
\,[FeIII]$3d^6G_4-3d^6H_4$      & 2.242 & 0.7 & 0.6 & 0.6 & 0.4 \\
\,H$_2$(2,1)S(1)                & 2.248 & 0.4 & 1.2 & 4.1 & 3.8 \\
\,??+H$_2$(3,2)S(2)             & 2.286 & 0.7 & 0.6 & 1.0 & 1.4 \\
\hline\hline
\end{tabular}}
\end{center}
Note: \hskip 0.1cm Intensity 100 corresponds to: \\
\hskip 1cm A~)  $2.65\times 10^{-3}$ erg cm$^{-2}$s$^{-1}$ sr$^{-1}$ \\
\hskip 1cm B~)  $2.08\times 10^{-3}$ erg cm$^{-2}$s$^{-1}$ sr$^{-1}$ \\
\hskip 1cm C~)  $0.61\times 10^{-3}$ erg cm$^{-2}$s$^{-1}$ sr$^{-1}$ \\
\hskip 1cm CS)  $0.50\times 10^{-3}$ erg cm$^{-2}$s$^{-1}$ sr$^{-1}$ \\
\end{table}

\section{Observations}

\subsection{ARNICA Observations}
The Orion Bar was 
observed during two observing runs in January 1996 and February 1997
using ARNICA (ARcetri Near Infrared CAmera)
mounted on the 1.5m TIRGO\footnote{The
  TIRGO telescope is operated by the C.A.I.S.M.I.-C.N.R
   Firenze, Italy} telescope. ARNICA is equipped with a 256x256 NICMOS3
array, the pixel size with the optics used 
at TIRGO is $0.96^{\prime\prime}$; for a complete
description of the instrument and of its performance, see Lisi et
al.~(\cite{Lea96}) and Hunt et al.~(\cite{Hea96}).
The Bar was imaged in the three J, H, and K broad band filters
(centered at 1.25, 1.65, and 2.2~$\mu$m, respectively) and in the
Br$\gamma$ narrow band filter ($\lambda=2.166\,\,\mu$m,
$\Delta\lambda/\lambda\sim 1\%$, Vanzi et al.~\cite{VGCT97}). 
The seeing was approximately 2-3\arcsec \  and
the observed field was approximately 
$\sim 4^\prime .5\times 4^\prime .5$,
covering all the Bar region.
Data reduction was carried out 
using the IRAF
\footnote{IRAF is made available to the astronomical
community by the National Optical Astronomy Observatories,
which are operated by AURA, Inc., under contract with the U.S.
National Science Foundation} and ARNICA
\footnote{A description of ARNICA can be obtained from the 
Arcetri Observatory at
{\tt ftp://150.217.20.1/pub/arnica/}} (Hunt et al.~\cite{HTBMM94}) software
packages. 
Photometric calibration in the J, H and K bands was
performed by observing photometric standard
stars from the list of Hunt et al.~(\cite{Hea97}) 
; the calibration
accuracy is estimated to be $\sim 5\%$.
The Br$\gamma$ image was continuum subtracted and calibrated using the
K band image.
We show in 
Fig.~\ref{fima} (left panel) an image obtained combining the  
J, H, and K images 
and (right panel) the continuum subtracted Br$\gamma$ image.

\subsection{LONGSP Observations} 

J (1.25\mic ), H(1.65 \mic ), and K (2.2\mic )
band spectra of the Orion Bar were obtained using the
LonGSp (Longslit Gornergrat Spectrometer) spectrometer mounted
at the Cassegrain focus on the TIRGO telescope.
The spectrometer is equipped with cooled reflective optics and
a grating in Littrow configuration. The detector is a
256$\times$256 engineering grade 
NICMOS3 array (for detector performances see Vanzi et al.~\cite{VMG95}).
The pixel sizes are 11.5 \AA\ (first order) and 1\farcs73 
in the dispersion and slit directions, respectively.
LONGSP operates in the range 0.9-2.5 \mic\ achieving
a spectral resolution at first order of $R\simeq550$ in J, 700 in H
and 950 in K.
For a more comprehensive description of the
instrument, refer to Vanzi et al.~(\cite{Vea97}).

Observations were conducted in two  runs 
in January and March 1996 under non-photometric
conditions. The slit used had dimensions
3\farcs5$\times$70\arcsec\ and was oriented N-S. The seeing
during the observations was in the range 2\arcsec-4\arcsec.
The Orion Bar was observed at three slit positions labeled
as 1, 2, 3 and shown in Fig.~\ref{fima} superimposed on a NIR
image obtained by co-adding the J, H and K 
ARNICA observations discussed in the previous section.
Position 1 and 2 were chosen in order to study the variation
of line intensities along a cut encompassing all the bar.
Position 3 was subsequently chosen to be coincident 
with the CS peak discovered by van der Werf et al.
at R.A.= 5$^h$ 32$^m$ 58.5$^s$ and Dec.= -5$^\circ$ 26$^\prime$ 25\arcsec\
(B1950.0).
This high density ($10^6$ cm$^{-3}$) clump appears to be
illuminated directly from the Trapezium and we thought it useful to examine
directly the relative variations in line
intensities across the clump. The center of the slits
were offset by -35\arcsec, -23\arcsec\ (Pos. 1) -35\arcsec, 23\arcsec (Pos. 2)
and 42\arcsec, 26\arcsec\ (Pos. 3) in R.A. and Dec. with respect to the
star $\theta^2$ A Ori. 

At each grating position we 
performed 5 ABBA cycles (A=on source, B=on sky) with an on-chip
integration time of 60 sec, for a total of 10 min integration on source.
At the beginning or at the end of the five cycles on the object
we performed 1 ABBA cycle on the O6 star BS 1895 ($\Theta^1$C Ori).

Data reduction was performed with the ESO package MIDAS, within
the context IRSPEC, modified to take into account LonGSp 
instrumental characteristics.
The frames were corrected for bad pixels,
flat-fielded, sky subtracted and wavelength
calibrated using the OH sky lines present
on all the frames (Oliva \& Origlia~\cite{OO92}).
After a direct subtraction, sky removal
was optimized by minimizing the standard deviation
in selected areas where the OH sky lines were poorly
subtracted but no object emission was present. 
The wavelength calibration was performed to better than 1/5 of
a pixel ($\simeq$2\AA).

The spectra were then corrected for telluric absorption 
by dividing by the  featureless spectrum of \THEC .
For more details on LonGSp data reduction, see Vanzi et al. \cite{Vea97}.

Flux calibration of the spectra in the J, H, and K bands
was achieved by rescaling the
observed flux distribution along the slit to match that obtained
from the ARNICA images at the positions of the slits.
We consider such calibration accurate to $\simeq$20\%
when comparing the fluxes of lines measured in two different bands.
Indeed, the comparison between the flux distributions
of H$_2$(1,0)S(1) $\lambda$2.12\mic\ in our observations
and in those by van der Werf et al. shows only a 10\% discrepancy
in the absolute flux level.

\subsection{IRSPEC Observations}
IRSPEC (Moorwood et al. \cite{Mea91})
observations of the bar using the ESO NTT telescope were carried
out in 1991 by Dr A. Moorwood. The detector was a SBRC 62x58 InSb array
with pixels of $\simeq$5\AA\ (H band )
along the dispersion
and 2\farcs2 along the slit direction.
The slit, 4\farcs4$\times$120\arcsec\ in size,
was oriented NE-SW as shown in
Fig~\ref{fima}. 
The data were uncalibrated and in this paper, we
merely make use of the profiles of line intensity along the slit. 

\begin{figure*}
\psfig{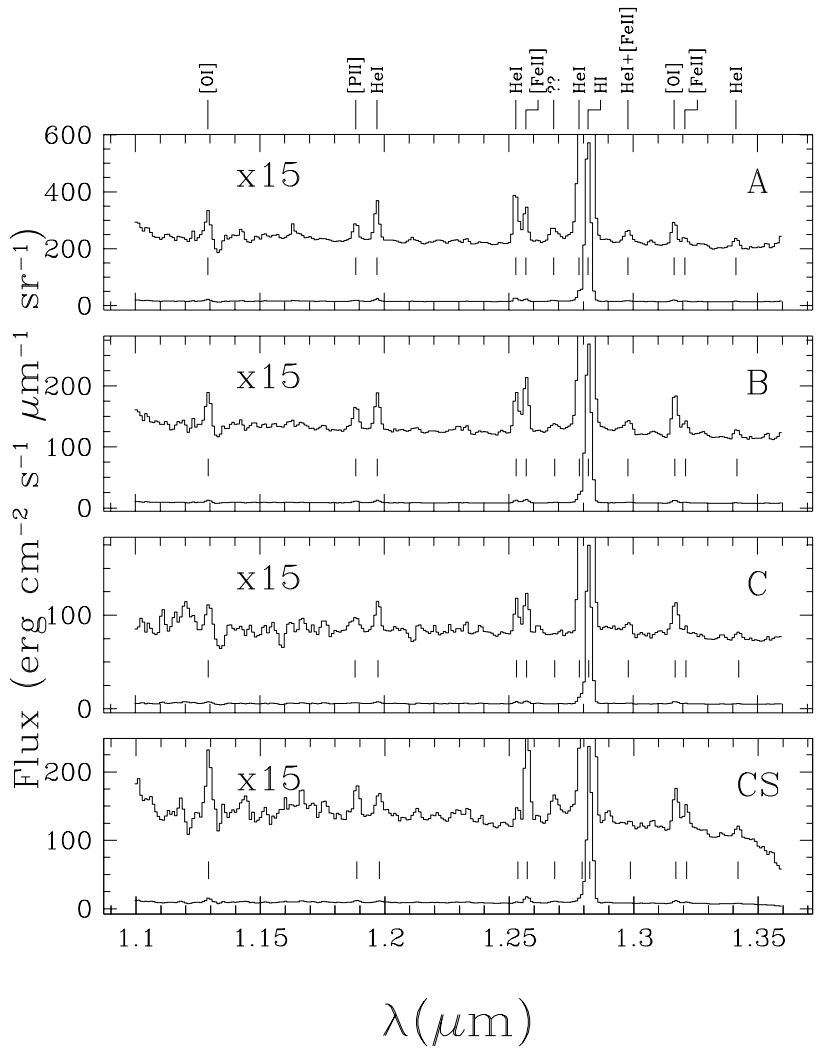}
\caption {  J band spectrum 
obtained towards the  
   three slit sections {\it A}, {\it B} and {\it C} and in the {\it CS}
position
  shown in Fig.1.
  The upper spectrum in each panel has been multiplied by 15 to emphasize 
  weak features.}
\label{fspecj}
\end{figure*}
\begin{figure*}
\psfig{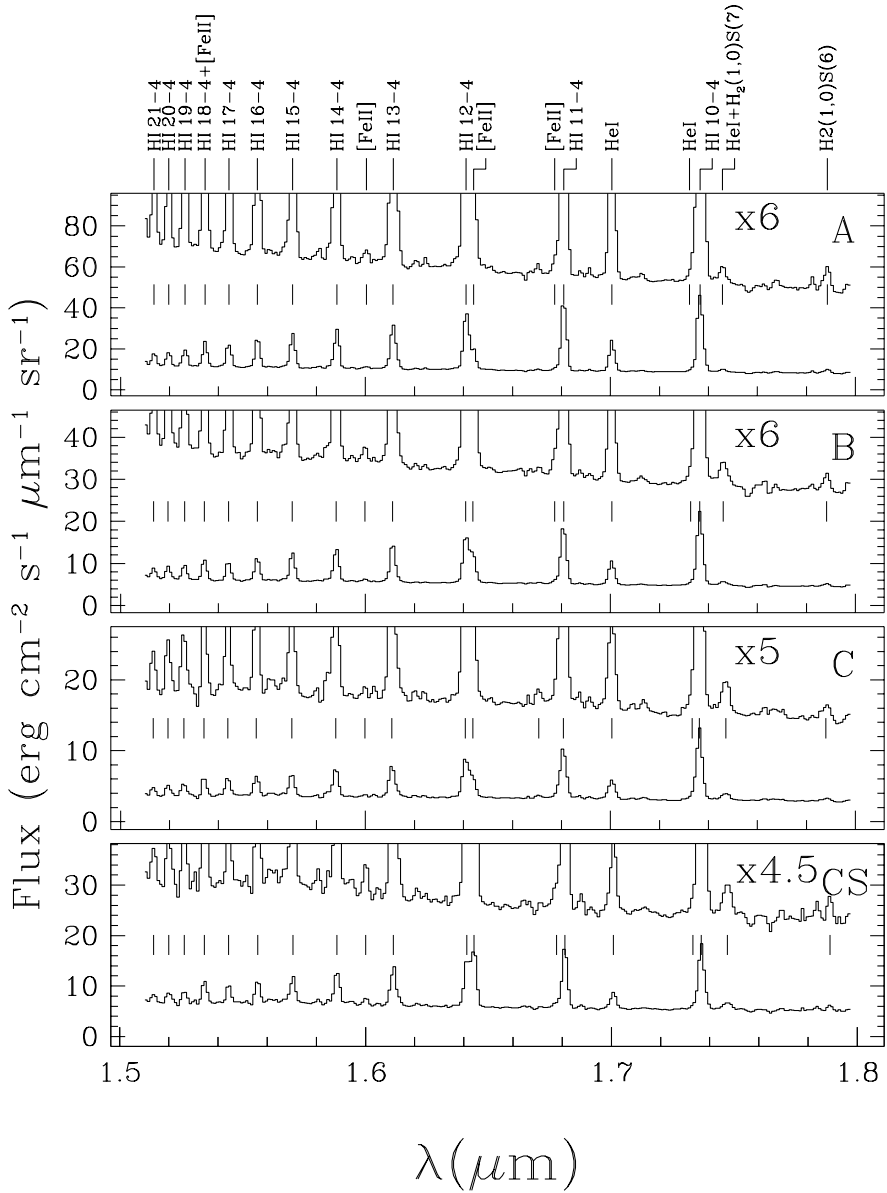}
\caption {  H band spectrum.} 
\label{fspech}
\end{figure*}
\begin{figure*}
\psfig{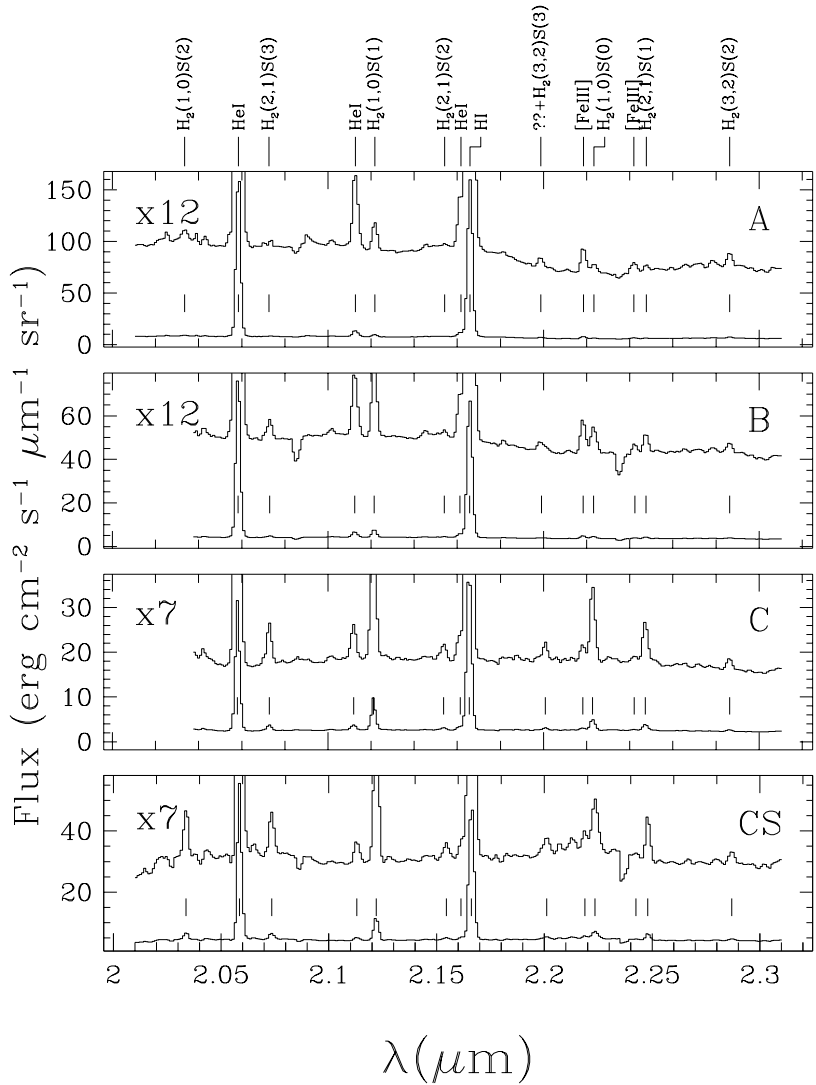}
\caption {  K band spectrum.}
\label{fspeck}
\end{figure*}

\section{Results} 

  In Figs.~\ref{fspecj}, ~\ref{fspech}, ~\ref{fspeck},
  we show sample spectra averaged over three portions of
  slit positions 1 and 2 (see Fig.~\ref{fima}). 
  Based on the profiles of line
  intensity along the N-S direction,
  we  decided to divide the  combined slit into three sections which
  we named  {\it A}
  (a 28 pixel section to the north),  {\it B} 
  (14 pixels in a central region),
  and  {\it C} (28 pixels to the south).
  These sections are displayed in the left panel
  of Fig. 1.
The hydrogen and helium recombination lines peak in position {\it A}
and
  become weaker to the south whereas the molecular hydrogen lines become
  stronger and reach their maximum intensity
in position {\it C}.  However,  there is clearly ionized gas towards the
region  {\it C} and vice-versa.  In Figs.~\ref{fspecj},~\ref{fspech},
  ~\ref{fspeck}, we show also
the spectra
  summed over 40 pixels in slit position 3 (the ``CS peak''). 

In Table 1, we give intensities corresponding to the
spectra shown in Figs.~\ref{fspecj},~\ref{fspech}~\ref{fspeck}. 
They have been averaged over the respective apertures.
Line intensities, in each region, are given relative to Br$\gamma$
(put equal to 100). 
Typical uncertainties vary from $\simeq10\%$ for strong lines
($I\simgreat 10$) to
$\simeq20\%$ for lines with $1\simless I\simless 10$
and about
50\% for the others. When comparing lines in two different
bands, a 20\% error due to spectrophotometric calibration
must also be taken into account.

The line intensities can be corrected for reddening using the Cardelli
et al. (1989) prescription A$_\lambda$/A$_V=0.48 \lambda^{-1.61}_\mu$,
where we have adopted $R$=5.5, as appropriate for the Orion region.
>From the ratio Pa$\beta$/Br$\gamma$ we derive A$_V\sim$ 2 mag (Sect.3.2).
Note that this value applies only to lines forming in the ionized 
gas.

\begin{figure}
{\psfig{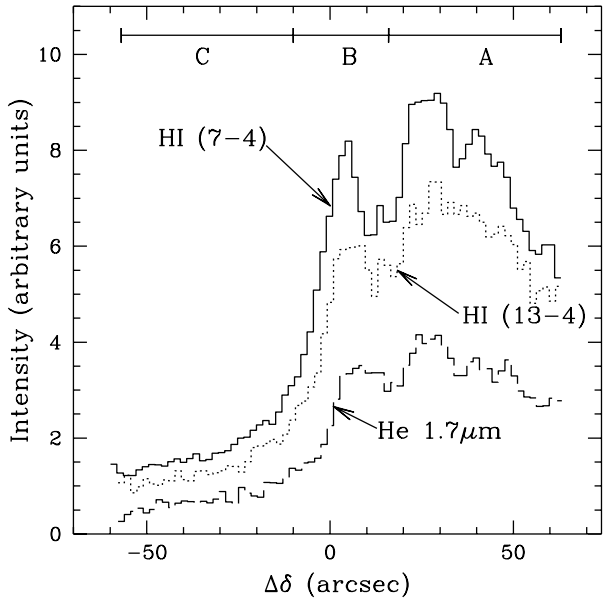}}
 \caption
  { Variation with declination offset (relative to declination (1950)
  =$-05^{\circ}$ 24\arcmin \ 51\arcsec ) of line intensities measured
  in the amalgamation of slit positions 1 and 2. The lines are
 H (7-4) (Br$\gamma $; solid) ,   H 13-4 (dotted), and the He line
  at 1.701 \mic ~(dashed). The vertical scale is arbitrary. We note that
  regions {\it A}, {\it B}, and {\it C} discussed in the text are
  defined as follows : {\it A}, $\Delta \delta > 16\arcsec $, {\it B},
  $-10\arcsec \, <\Delta \delta \ < \, 16\arcsec $, {\it C},
  $\Delta \delta <\, -10$\arcsec .}
\label{fcuthi}
\end{figure}

\begin{figure}
{\psfig{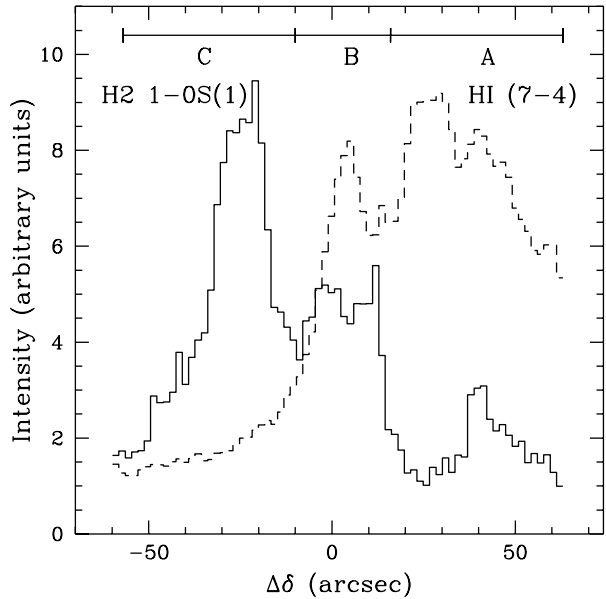}}
 \caption
{ Variation with declination offset of H (7-4) (dashed line) and
 \MOLH (1-0)S(1) (solid line), 
 measured
 in the amalgamation of slit positions 1 and 2.  The regions defined
 as {\it A}, {\it B}, and {\it C} are shown.}
\label{fcuth2}
\end{figure}

\begin{figure}
{\psfig{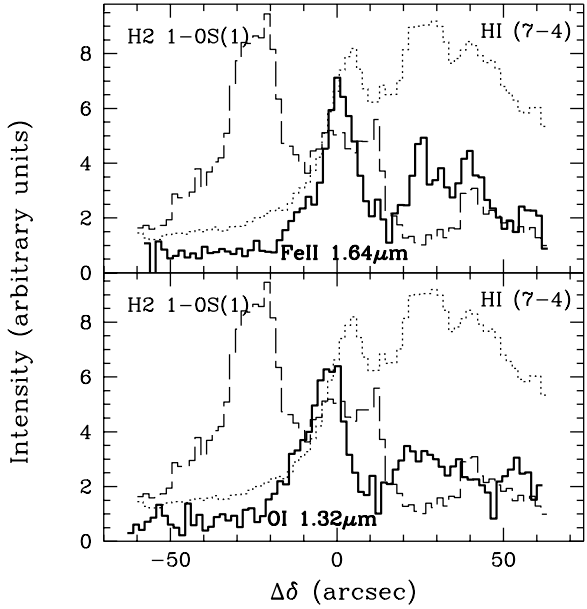}}
 \caption
{ Top panel: variation with declination offset of H (7-4) (dotted line),
H2 1-0S(1) (dashed line)
and FeII 1.644 \um (solid line). 
Bottom panel: variation with declination offset of H (7-4) (dotted line),
H2 1-0S(1) (dashed line)
and OI 1.317 \um (solid line). 
In both cases, the variations are
 measured
 in the amalgamation of slit positions 1 and 2. }
\label{fcutfeoi}
\end{figure}

\begin{figure}
{\psfig{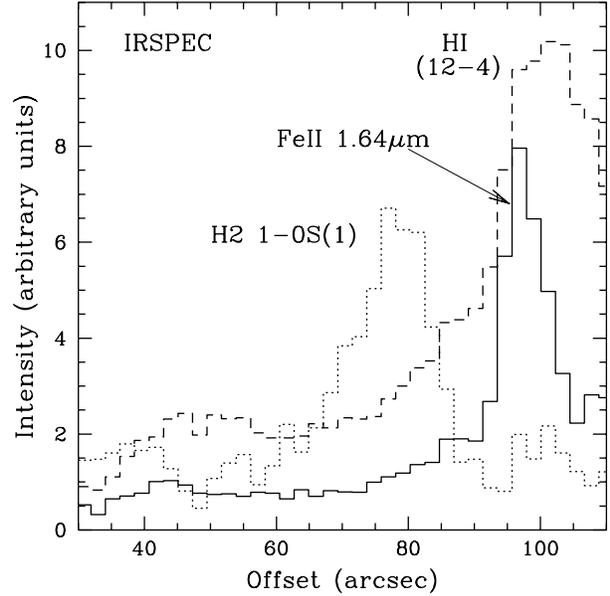}}
\caption
  { IRSPEC cuts: HI (12-4) (dashed line), H2 1-0S(1) (dotted line)
and FeII 1.644 \um (solid line).}
\label{fcutirspec}
\end{figure}

\begin{figure}
{\psfig{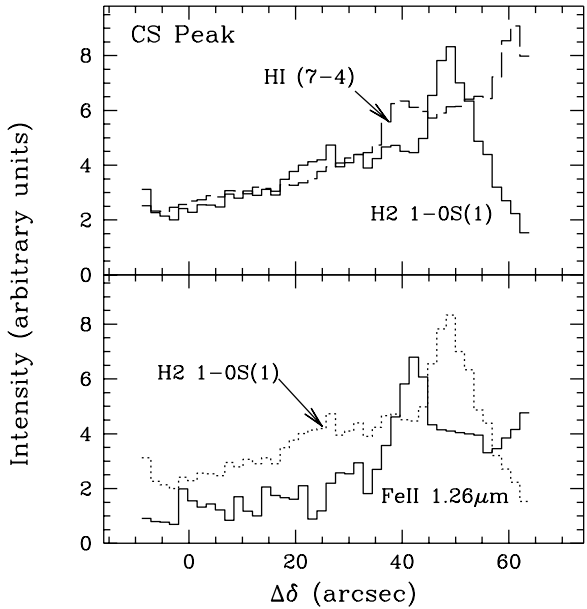}}
\caption
  {Variation with declination offset along the slit centered on the CS
peak of selected lines:  HI (7-4) (dashed line, top panel),
H2 1-0S(1) (solid line, top panel and dotted line, bottom panel),
and FeII 1.257 \um (solid line, bottom panel).}
\label{fcutcs}
\end{figure}

The variation in intensity along the 
amalgamation of slit positions 1 and 2 of a 
variety of interesting line tracers 
is shown in Figs.~\ref{fcuthi} to~\ref{fcutfeoi}. 
Figure ~\ref{fcutirspec} shows intensity variations 
in a number of lines from the IRSPEC
data. Figure ~\ref{fcutcs} plots the intensity profile of selected
lines at slit position 3.
>From Figs. ~\ref{fcuthi}-\ref{fcutfeoi},
we can see that  region {\it A} coincides roughly with a peak in the
lines of HI, which have a second peak in {\it B},
while the emission of molecular hydrogen has a strong
peak in {\it C} and a weaker one in {\it B}. Lines of FeII and OI
have a sharp peak in {\it B} and a secondary one in {\it A}, but
are absent in region {\it C}. In the following, we will use the
definition {\it A}, {\it B} and {\it C} 
(see caption to Fig.  ~\ref{fcuthi}) both to identify  the
peaks in the intensity profiles and  with reference to Table 1.

We now summarize our results starting with the IRSPEC data which have
the advantage that the slit is oriented perpendicularly to the bar.
We then discuss in turn the hydrogen and helium recombination lines
(which we presume form in the ionized gas), the two oxygen
transitions, the collisionally excited
iron lines which may form close to the ionization front, 
and the molecular hydrogen lines which 
are thought to form in hot neutral gas close to the ionization front.

%

\subsection{IRSPEC cut}
 The observations made using IRSPEC provide a useful introduction to the
 TIRGO results which have a wider spectral coverage. Figure
 ~\ref{fcutirspec} compares profiles in the Br12 line from the
 ionized gas, in the FeII 1.644\mic \ ($4sa^{4}$D$_{7/2}$-$3d^7a^{4}$F$_{9/2}$)
 line which traces gas close to the ionization front (see below), and
 in the molecular hydrogen v=1-0 S(1) line from hot (T$>$ 1000K)
 molecular gas. Figure ~\ref{fcutirspec} demonstrates the fact that 
 the molecular hydrogen peak is offset $\sim$16\arcsec (0.035 pc) from the
 ionization front as marked by FeII (or by the fall-off in Br12).
 The simplest explanation of this (see Tielens et al. 1993, van der
 Werf et al. 1996) is that one is observing an edge-on PDR  and that 
 the observed offset corresponds to the difference between the
 ionization front where Lyman continuum photons are absorbed and the
 \MOLH \ dissociation front where photons capable of dissociating
 molecular hydrogen are absorbed.  We note also that an offset of
 16\arcsec \ in the IRSPEC data  corresponds (given the orientation
 NE-SW of the bar) to 23\arcsec\ in
 the TIRGO slit oriented N-S. 
\subsection {HI lines}

In the LONGSP data,
we detect  many recombination lines of atomic hydrogen, 13 lines of
the Brackett series (Br$\gamma$ in the K band and 12 lines, from
(10-4) to (21-4) in the H band), and  
Pa$\beta$ and Pa$\gamma$ in the J band.

\begin{figure}
{\psfig{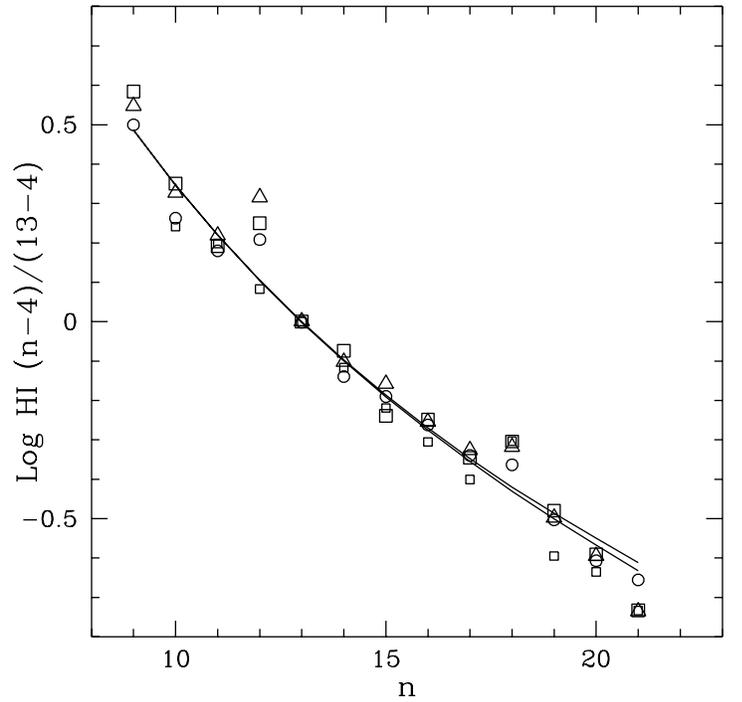}}
\caption{Ratio of the H-band hydrogen recombination lines (n-4)/(13-4) as a function
of the quantum number n. Filled circles refer to position {\it A},
triangles to {\it B}, squares to {\it C}, and diamonds to 
position CS (the CS peak). 
The  line
(18-4) is blended with an [FeII] line.
The solid line shows the predictions of recombination theory  (Case B).
The dashed line shows
the Case B ratios corrected for a reddening of \Av=2 mag.}
\label{fhi}
\end{figure}

We use the 13 lines in the Brackett series to check the accuracy of our data.
Figure ~\ref{fhi} plots the ratio (n-4)/(13-4) as a function of the quantum number n
and compares our results with the prediction of recombination theory
(Storey \& Hummer 1995). The agreement is quite good, 
well within our estimate of 10-20\% for the 
observational errors. 

The ratio of Pa$\beta$/Br$\gamma$ provides a value of the extinction
\Av$\sim$2 mag, 
assuming the reddening
curve of Cardelli et al. (1989) and $R$=5.5.
We detect a slight variation along slit 1+2, from 2.3 mag in 
{\it A} to 1.4 mag in {\it C}. This variation, however, is within
our estimated 25\% error in this line ratio.
The value in the CS position is \Av=1.6 mag.
Figure ~\ref{fhi} shows as a dashed line the theoretical line ratios
corrected for \Av= 2 mag.

\subsection {He lines}
 One of the aims of our observations was to examine the extent to which 
 helium is neutral within the zone of ionized gas.  Estimates of the helium abundance
 based upon measurements of either radio or optical recombination lines often
 assume that the helium and hydrogen Str\"{o}mgren spheres are coincident with
 one another (see e.g. Mezger 1980).  Our profiles along the slit allow us
 to make a direct comparison of the HeI and HI line intensities which 
 can then in principle be transformed into the abundance ratio 
 $[He^{+}]/[H^{+}]$ in the immediate vicinity of the ionization front
 of the Bar.  
 
 The chief obstacle in doing this is the  uncertainty in helium line 
 intensities which results from collisional excitations from  the 
 metastable 2$^{3}$S and 2$^{1}$S states.  Smits (1996) has computed
 helium line intensities in an approximation where collisions (and
 self-absorption) out of the metastable levels  into n=3 and 4 are neglected
 although collisions between the n=2 levels are considered.  We have compared 
 our observed intensities with Smits predictions for electron density
 $10^4$ \percc \ and temperature $10^4$ K.  We normalise for this purpose to
 the 1.701\mic\ 4$^{3}$D-3$^{3}$P$^\circ$
 transition which has the same upper level as
 the 4471 \AA \ line often used in optical analyses.  It is 
 expected that this transition (see Osterbrock et al. 1992) is only
 affected at the 1-2 percent level by the collisional effects mentioned
 above and we neglect such effects in the following. One finds then 
  that, relative to 4$^{3}$D-3$^{3}$P$^\circ$, lines such as
  5$^{3}$D-3$^{3}$P are in good agreement with the Smits
  predictions but 4$^{1}$S-3$^{1}$P$^\circ$ and 5$^{1}$S-3$^{1}$P at
 2.113 and 1.341\mic  \ respectively are factors of roughly 3 stronger.
 The reason for this  may be  the neglect of the
 collisional effects and trapping discussed above (see e.g. Robbins
\& Bernat 1973; Peimbert \& Torres-Peimbert 1977). 
   We  in any case have assumed 
 that the 1.701 \mic \
 line    behaves essentially as predicted by the Smits
 models and can be used to estimate the He$^{+}$ abundance. 
 It is natural to compare the He 1.701\mic \ intensity with
 that of the adjacent Br10 line. With the above assumptions, we
 find that :
 \begin{equation}
  F(He,1.7\mu )/F(H,10-4) \, = \, 3.61 \, \frac{[He^{+}]}{[H^{+}]}
  \end{equation}

\begin{figure}
{\psfig{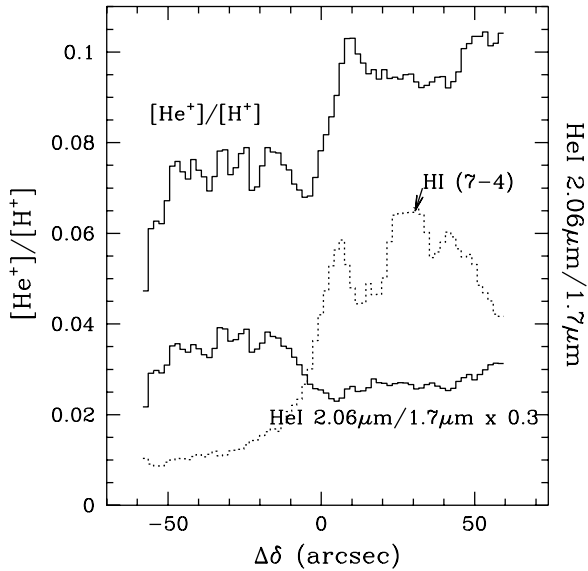}}
\caption{The figure shows the He abundance [He$^+$]/[H$^+$] and
the ratio of the two HeI lines at 2.058 and 1.701 \um\ as a
function of the declination along the amalgamation of slits 1+2,
compared with the Br$\gamma$ profile.
The data have been smoothed over three pixels. }
\label{fhehi}
\end{figure}

 In Fig.~\ref{fhehi}, we show the profiles along the slit of the  
  abundance ratio $[He^{+}]/[H^{+}]$ 
 deduced from Eq.(1) as well as the ratio of the 2.06 
 (2$^{1}$P$^\circ$-2$^{1}$S) to the 1.7\mic \ lines. The Br$\gamma \ $
 profile along the slit is shown for comparison. 
 We derive an abundance ratio of $0.093\pm 0.005$  over region {\it A}
 consistent with He abundance estimates from other authors
 (see e.g. Baldwin et al.1991 who find 0.088 based on optical
 measurements and the review of Mezger 1980 (his Fig.5) who 
 shows that radio estimates at positions less than 100\arcsec \
 from \THEC \ are in the range 0.083-0.09).
 It appears  that at the position of our slit, the helium and hydrogen
 Str\"{o}mgren spheres are close to being coincident.  
 We note with interest the decrease of
 $[He^{+}]/[H^{+}]$ to values of $\sim 0.075$
 in the southern part of the slit or effectively in 
 zone {\it C}, where  
we seem to observe different ionization conditions in this (presumably)
foreground ionized material.
 One notes also
 the  fact that the degree of ``enhancement'' of the 2.06
 micron line seems to increase slightly (by a factor of 1.3) in 
 zone {\it C}. 

\subsection {OI lines}

We detect two OI lines in the J band, the \OIa\ at 1.129\um\
and the \OIb\ at 1.317\um, with comparable intensity.
These lines are produced in the neutral gas by  
excitation of OI to the upper level
of the transitions by UV photons, at 1027 and 1040 \AA\ respectively,
followed by radiative decay.  
The fact that the two lines have similar intensity
(though not at the CS peak, see Fig.2)
suggests that the contribution of Ly$-\beta$ photons to the excitation
of the 2p$^3$3D level  (the upper level
of the 1027 \AA\ line) is negligible.
That the excitation mechanism is fluorescence 
is confirmed by the spatial
variation of the line intensities, 
which peak at the edge of the ionized region
as marked by the HI recombination
line emission (cf. Fig.~\ref{fcutfeoi}).
In other words, the oxygen lines are an excellent marker
of the ionization front. More precisely, one
can say that
due to the rapid charge exchange of OI with
H$^{+}$ and the fact that the hydrogen and oxygen ionization 
potentials are close to being identical,  the oxygen  
lines trace neutral gas close to the ionization front. 

It is possible 
to use the OI lines to derive a measure of the UV radiation
field G$_0$ at the edge of the bar. If fluorescence is the
dominant excitation mechanism,
the number of photons emitted in the IR line equals the number of 
photons absorbed by the UV line. Since the optical depth of the UV
lines is always very large ($\tau_0\sim 5 \,N/10^{18}$  for the 1040 \AA\
line and $\sim 10\,N/10^{18}$ fo the 1027 \AA\,
where $N$ is the atomic H column density and we have assumed
a velocity dispersion $\Delta v$=3 km s$^{-1}$ 
and O/H=6$\times 10^{-4}$), the number of absorbed UV photons is
proportional to the line equivalent width in the ``flat'' portion of
the curve of growth.  This is given by :
 $W_\lambda/\lambda_{UV}\sim 1.2\Delta v/c \>F(\tau_0) \sim 3.6\times10^{-5}$
(Spitzer 1978, p53, where $\tau _0$ is the UV line center optical depth
and F($\tau _0) \sim 3$ for optical depths of order 1000). 
Since both UV lines are in fact triplets with separation
larger than $W_\lambda$, the total number of UV photons absorbed 
and re-emitted in each IR line is given by:

\begin{equation}
I_\nu^{UV}= {{4\pi \sin \theta_t}\over{3}}{{\lambda_{IR}\lambda_{UV}}\over{c W_\lambda}}\>\>
I(IR)\>\>\>\> {\rm (erg\, cm^{-2} s^{-1} Hz^{-1})} 
\end{equation}
where I(IR) is the observed intensity of the IR line in 
erg cm$^{-2}$ s$^{-1}$ sr$^{-1}$ and $\theta_t$ is the angle between
the  PDR and the line of sight ($\theta_t=90\deg$ in a face-on PDR;
see Appendix). 
Note that the three components of the IR lines are not resolved in our spectra.

Equation 2 gives   $I_\nu^{UV} \sim 1.2\times 10^{-13}\sin{\theta_t}$ for
$I(IR)=2.6\times 10^{-4}$ erg cm$^{-2}$ s$^{-1}$ sr$^{-1}$,
as observed in the main peak of the 1.317 \um\ line.
Here, we have corrected for 2 magnitudes of visual 
extinction (see Sect. 3.2).
The inferred UV intensity can be compared to the
flux from \THEC \ at the projected distance 
of the bar ($I_{p}\, \sim 4\times 10^{-14}$
erg cm${-2}$ s$^{-1}$ Hz$^{-1}$   for
\Tstar=40000K, \Lstar=2.5$\times 10^5$ \Lsun). The
main uncertainty is the appropriate value for $\theta_t$ \
but  the Orion Bar is known to be close to edge on
(see e.g Hogerheijde et al. 1995). A plausible
value is $\sin\theta_t\sim 0.2$ ($\theta_t\sim10-15 \deg$;
see Sect. 3.6), and one finds then 
 that the physical distance
of the bar from \THEC\ is about 
$(I_{p}/I_\nu^{UV})^{0.5}$ or 1.3 times the projected distance. 

The UV intensity can be expressed  relative to
the interstellar diffuse field  taken here to be 
 $3\times 10^7$
photons cm$^{-2}$ s$^{-1}$. The normalized UV intensity is
then $1.3\times 10^{5}\, \sin {\theta_t}$ or 
G$_0\sim 2.6\times 10^4$ (for $\sin\theta_t$=0.2), similar to the value used by Tielens et al. (1993).

The second, weaker peak of emission {\it A} at $\Delta\delta\sim$23 arcsec
has an OI intensity about two times lower than the main peak.
This may be  due to a different orientation of the front
with respect to the line of sight. 
The OI intensity on the CS peak is about 1/4 that of the main peak
which again may be due to an orientation effect
although it could also imply  dust extinction between the CS peak and the
Trapezium. In general, our OI results imply G$_{0}$ values at the
ionization front in the range $6000 - 3\times 10^4$.

\subsection{Iron lines}

  Model calculations suggest  (see Baldwin et al. 1991, Rubin et al.
1991, Osterbrock et al. 1992)  that iron is mainly in the form FeIV 
in the Orion nebula.  It follows that one expects to see FeII and
FeIII emission predominantly at the edge of the Str\"{o}mgren  sphere
close to the ionization front.  In Fig.~\ref{fcutfeoi}, we show profiles 
along our amalgamated slit of the 1.644 \mic \ [FeII]
 transition compared with the
 1.317\mic \ OI line discussed in  the previous section.
 One sees that
 the two lines show rather similar behavior  with a peak 
 slightly offset from
 one of the maxima observed in the H recombination lines. 
 It seems plausible that this approximate coincidence denotes the
 presence of an ionization front and the OI data which we discussed above
 confirm this idea. It is notable also that the iron lines show no
 evidence for a coincidence with the molecular hydrogen 1-0 S(1)
 {\it C} peak at $\Delta \delta $= -27\arcsec \
 which is also shown on Fig.~\ref{fcutfeoi}.
 In fact,
  our data suggest that both the FeII and FeIII emission lines form
  in ionized (or partially ionized but not neutral)
  gas close to the ionization front. It is worth noting here
  that extinction estimates which we have made using the ratio of the
  1.644 $4sa^{4}$D$_{7/2}$-$3d^7a^{4}$F$_{9/2}$ to the 1.26\mic \
  $4sa^{4}$D$_{7/2}$-$4sa^{6}$D$_{9/2}$ transitions is consistent with
  a visual extinction of $2.7\pm 0.9$ magnitudes at all positions.

 We can estimate the conditions required to
 explain the relative intensities of the FeII lines.  FeII
 line ratios  can be used as indicators of
 electron density (see Oliva et al. 1990, Pradhan \& Zhang 1993).
  Our most useful indicator appears to be the ratio of the
  1.600 $4sa^{4}$D$_{3/2}$-$3d^7a^{4}$F$_{7/2}$ line intensity
  to that of the 1.644 $4sa^{4}$D$_{7/2}$-$3d^7a^{4}$F$_{9/2}$. 
  We find that this ratio varies in the range 0.06-0.1 
  (see table 1) over the regions covered by our slit.
  Based on the collisional rates of Pradhan \& Zhang(1993) 
  (see also Oliva et al. 1990), we conclude that this
  corresponds to an electron density  n$_{e}$  of 4000 \percc \
  in the  region {\it A}), 6000 \percc  \ in
region {\it C}, and 3000 \percc \ in region {\it B}.
  At the CS peak position, the observed 1.600/1.644 ratio
  is 0.12 corresponding to n$_{e} = 10^4 $ \percc .
   Thus, we can exclude high density clumps 
  with n$_{e}$ = $10^6$ \percc \
  of the type discussed by Bautista et al.(1994).  On the
  contrary, the FeII data seem consistent with ionized gas of
  electron 
  density $\sim $ \ $10^4$ \percc \ or less 
  in the vicinity of the ionization
  front.  It is worth stressing here that this line  ratio converges to the
  LTE value at electron
  densities of roughly $10^5$ \percc and so the lack of
  high density clumps is significant. 
  Moreover, the ionization degree in the layer where the FeII lines are
  formed is unlikely to be much smaller than unity and hence the
  hydrogen density is also likely to be of order 10$^4$ \percc.

\subsection {H$_2$ lines}

\begin{figure}
{\psfig{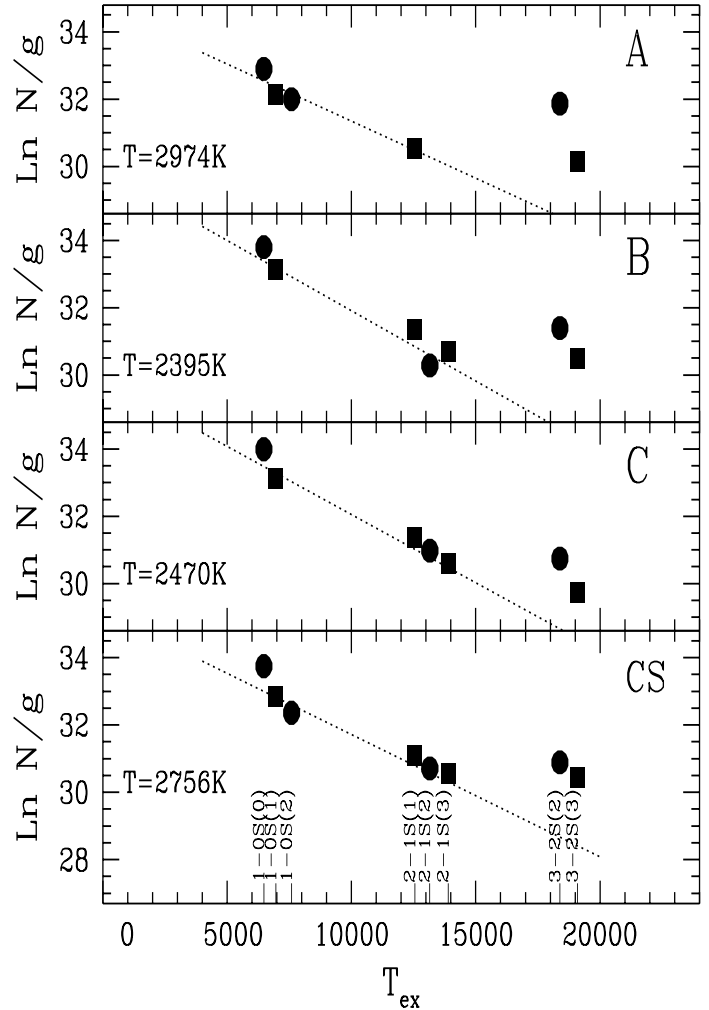}}
 \caption
{Column density per sub--level against excitation energy of the level
  at positions
{\it A, B}, {\it C} and on the {\it CS} Peak.
The dotted lines show the best fit to a single temperature
population. The two v=3 transitions are doubtful and have not been
considered on the fit.
 Filled squares  represent ortho transitions  and
 circles  para transitions.}
\label{fh2}
\end{figure}

We measured the intensity of 8 H$_2$ lines in the K band,
covering an excitation range  from 6472 to 
18089 K (see Table 1).  Using the data from table 1 and
transition probabilities from Turner et al. (1977), we
have determined upper level column densities  at positions
{\it A, B}, {\it C} and {\it CS}. 
In Fig.~\ref{fh2}, we  plot the column densities
per sub--level derived in this manner against excitation energy. 
One sees that there are clear departures from an LTE distribution
suggesting either  that fluorescence is playing a role in determining
level populations or that there is a sharp gradient of temperature along
the lines of sight sampled.  The ``best fit temperatures" derived from
fitting a Boltzmann population distribution to the data in Fig.~\ref{fh2}
are moreover rather high with values ranging from $\sim$2500 K in 
region {\it C} to  3000 K in region {\it A}. 

\begin{figure}
{\psfig{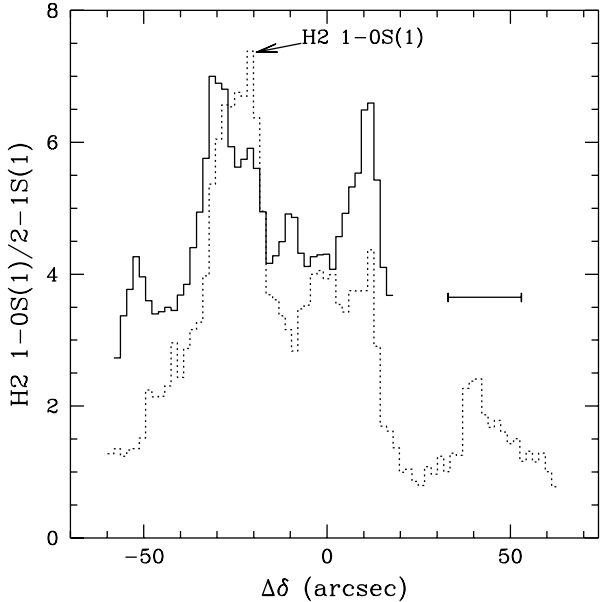}}
 \caption
{Ratio of the 1-0S(1) to 2-1S(1) H$_2$ lines as a function
of the declination offset along our amalgamation
of slits 1+2
(solid line). To compute the ratio, the profiles have
been smoothed over three pixels. The horizontal bar shows the
value averaged over declination offset 0-50 arcsec.
 We show for comparison the profile of the 1-0S(1) line
(dotted line).}
\label{fhr}
\end{figure}

\begin{figure}
{\psfig{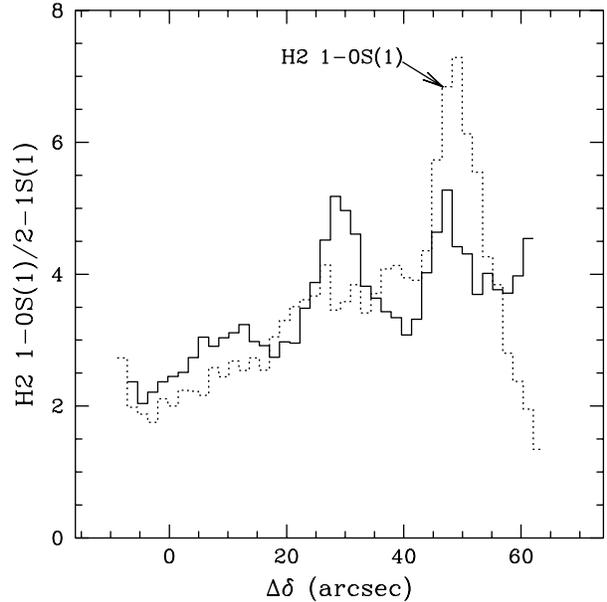}}
 \caption
{ Same as Fig.13 for the CS position.}
\label{fhrcs}
\end{figure}

The high excitation temperatures  as well as 
the (probable) detection of lines from levels as high as v=3 suggest  that
 we are detecting  extended
fluorescent emission (seen on larger scales by Luhman \& Jaffe, 1996)
in addition to
a ``thermal'' layer. 

The evidence that some fluorescent emission is present
 is strengthened by the behaviour of
the intensity ratio of  the v=1-0 S(1)
 (2.12\mic ) and v=2-1 S(1) (2.25 \mic ) \MOLH \ lines along our 
 amalgamated slit ({\it A,B,C}), shown in Fig.~\ref{fhr}.
 In a pure fluorescent model, this ratio is predicted to be
 $\sim$2, whereas an admixture of collisional excitation
 leads to higher values (12 for pure thermal emission
 at 2000 K).
  Our 
 results are consistent with those of van der Werf et al. (1996)
in that
 at the main molecular hydrogen intensity peak {\it C},
the ratio is $\sim$ 6-7.
A similar value is also found at the secondary peak {\it B},
which is about coincident with	
the position of the ionization front as traced by the OI lines. Over
 the rest of the slit
 the ratio is $\simless$4.
 Thus the fraction of fluorescent emission is smaller at the peaks
 of v=1-0 S(1) emission.

\begin{figure*}
{\psfig{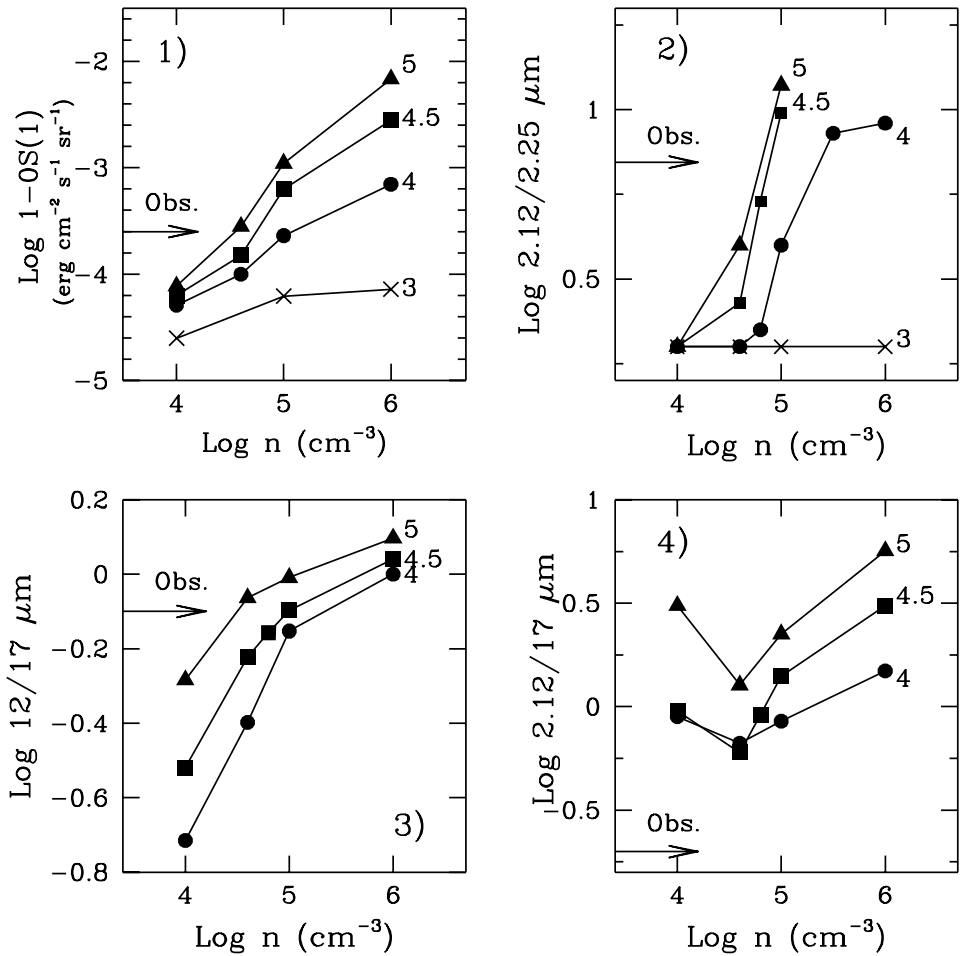}}
 \caption
{ Panel (1): model predictions for the intensity of the H$_2$ 1-0 S(1) line
as a function of hydrogen number density for different values of G$_0$. 
Panel (2):  ratio of the v=1-0 S(1) line at 2.12 \um\ to the v=2-1 S(1) line 
at 2.24 \um\ as a function of density.
Panel (3): ratio of the v=0,J=4-2 line at 12 \um\ to the v=0,J=3-1 at 17 \um.
Panel (4): ratio of the v=1-0 S(1) line to the v=0,J=4-2.
Each curve is labelled by the logarithm of G$_0$. Triangles indicate
G$_0=10^5$, squares with G$_0=10^{4.5}$, circles with G$_0=10^4$, and
crosses with G$_0=10^3$. The observed value at peak {\it C} is shown
by the arrow in each panel.}
\label{f4w}
\end{figure*}

These measurements, as well as 
the observed intensity of the v=1-0 S(1) line,
can be compared to the  PDR model calculations
of Hollenbach \& Natta (1995) (at steady state). Figure
 ~\ref{f4w} shows  the
predicted intensity of the 1-0 S(1) line in a face-on PDR
as a function of the
density for different values of G$_0$, and in Panel (2) 
the ratio of the 1-0 S(1) to the 2-1 S(1) line.
In the main molecular peak {\it C}, we observe a peak intensity of the 1-0 S(1) line
of $\sim 2.3\times 10^{-4}$ \sbu, and a ratio of $\sim$7. 
If we assume that the  H$_2$ lines are produced in a PDR with 
G$_0\sim 3\times 10^4$ (see Sect. 3.3), this corresponds to a density
$6\times 10^4$  cm$^{-3}$.

These  results  assume a face--on PDR and there is good
reason to suppose that the Orion Bar is seen edge on
(see Jansen et al. ~\cite{Jea95b}  for a discussion of the geometry). 
The  effect of a slant of the Bar on
the \MOLH \ lines is complicated by the effects of dust extinction
and a brief discussion is given in the appendix. 
The results are different for the vibrationally excited H$_2$ lines
at 2\um\ and for lines at longer wavelengths, for which the
extinction is negligible.
Parmar et al. (1991) have observed the J=3-1 (17\mic ) and
4-2 (12.3\mic ) v=0 \MOLH \ lines toward the bar and find that 
their observations
are compatible with a hydrogen column density of $3\times 10^{21}$ \cmsq 
\ at a temperature of order 500 K.
Figure ~\ref{f4w} shows in Panel (3) the model-predicted 
ratio for the two lines observed
by Parmar et al. and in Panel (4) the ratio of the v=2-1 S(1) to the
J=3-1 (17\mic) lines. These ratios can be well reproduced by a
model with G$_0\sim 3\times 10^4$ and $n\sim 6\times 10^4$ \percc\
having 
a moderate enhancement of the intensity of the 12 and 17 \mic\ lines
due to an inclination of the Bar with respect to the line of sight
$\theta_t\sim$10$\deg$. These values are also consistent with
the inclination required to explain the intensity of the OI lines, discussed in 
Sect. 3.4.

The density in the other regions where molecular lines are measured can be
estimated in a similar way from Fig. ~\ref{f4w}.
In the secondary molecular peak {\it B}, assuming the same inclination
$\theta_t\sim 10\deg$, we obtain $n\sim 4\times 10^4$ \percc.
 These densities are a factor of $\sim$5-10 larger than  that of the
ionized gas (see Sect. 3.5) and roughly consistent with the density
required 
to explain the stratification of the bar (i.e the offsets between
cold molecular gas and ionization front, see Tielens et al. 1993),
which is of order $5\times 10^4$ \percc .


 Similar densities can be derived from our
data at slit position 3 , the ``CS peak'' ({\it CS}). 
We show the 1-0 S(1) intensity in Fig.~\ref{fcutcs}
and the v=1-0/2-1 line ratio in Fig. ~\ref{fhrcs}.
We see  a peak of H$_2$ emission at $\Delta\delta\sim 49$\arcsec
with an intensity of $\sim 1.6\times 10^{-4}$\sbu,
and a broader emission between 0\arcsec and 30\arcsec, with a  peak intensity
of $\sim 8\times 10^{-5}$ \sbu. In both cases the ratio of the two H$_2$
lines is about
5 implying densities $\simless 10^5$ \percc \ for
G$_0\sim 10^4$.
Neither of the peaks coincide
with the ``ionization front'' as defined by FeII (Fig. ~\ref{fcutcs}).
One would expect such a coincidence if the densities were considerably
above $10^5$ \percc \ as implied by the CS data (see van der Werf et
al.1996).  Thus the general conclusion is that  the PDR as seen in
molecular hydrogen appears to be at densities below $10^5$ \percc \
 in contrast to the millimeter results which suggest
the existence of  clumps at densities well above $10^5$ \percc .
%
%
%

  A caveat to much of the above discussion is that comparisons
 which we have made between the predictions of the Hollenbach
\& Natta code used by us and other results (in particular,
the results of St\"{o}rzer \& Hollenbach 1997)  shows that there 
can be substantial differences in both the \MOLH \ line intensities
and in some line ratios. The main reason for this is the extreme
sensitivity of \MOLH \ line intensities to the temperature structure
although there are more minor effects which also play a role
(note for example 
that the Hollenbach Natta models neglect the effects of
the carbon oxygen chemistry upon \MOLH ). This has the consequence that
rather minor errors in the treatment of thermal equilibrium can
affect our conclusions.  In this respect, mid infrared 
intensity ratios such as
the 12/17 micron ratio measured by Parmar et al. are  important
 because they afford a direct measure of temperature. Thus if the    
temperature structure can be adjusted  to fit the mid infrared
observations, one may have some confidence in the predictions
for other lines. For the moment, we conclude that
 the fact that we have been able to fit
 the 12/17 ratio suggests that the model which we have used is
 reliable.

\section{Discussion}
Our analysis of  molecular
 hydrogen in the previous section 
 has assumed implicitly that the H$_2$ emission
comes at all positions from material heated by the stellar
radiation field (PDR). 
The two peaks {\it C} and {\it B} we observe must then come from two different structures.
In fact, there are
 indications in the molecular hydrogen images of van der Werf et al. that in
 \MOLH \ lines there are {\it two} bars.  
One of these (which one might
 call the main \MOLH \ bar)  corresponds to the feature seen in our
 Fig. ~\ref{fcuth2} at position {\it C} ($\Delta \delta = -24$\arcsec ).
The        
 ``second \MOLH \ bar'' (less well defined in the van der Werf et al.
 images) is close in projection to the main ionization front and corresponds
 to the peak in Fig. ~\ref{fcuth2} at  $\Delta \delta$=-2\arcsec \ 
 (roughly coincident with the OI peak).  We propose that these
 two bars are separated along the line of sight and that the shift in
 position (15\arcsec \ accounting for our slit orientation)
 is due to a tilt in the bar of $\sim$10 degrees as discussed above.
 Thus, the bar is split along its  length
 (essentially along the line of sight) into two sections separated by
 0.2-0.3 parsec.  Each half of the bar in this scenario has a length 
 of order 0.1 parsec and thus the total length along the line of
 sight is  of order 0.3-0.4 parsec or somewhat smaller
 than  proposed by Jansen et al.
 (~\cite{Jea95b}). 

 There are problems with this model however. The
 principal difficulty is that our
 data do not show evidence for ionization front indicators 
 (OI and FeII) roughly
 coincident (one expects a shift of order 6 arc second for a
 density of $5\, 10^4$ \percc ) with the main 
 ($\Delta \delta = -24$\arcsec \ on our slit) molecular hydrogen
 peak.   One explanation for this might be that the density in the
 layer of gas between the ionization and \MOLH \ dissociation front is 
 a factor of 2-3 lower (of order $2\times 10^4$ \percc \ )
 and thus that the main \MOLH \ peak is  shifted
 15\arcsec \ (20\arcsec in our NS oriented slit) relative to the main ionization
 front (i.e., peak {\it C} in \MOLH \ corresponds to peak
 {\it B} in OI ).  Equally, the ionization front emission seen in
 Fig. ~\ref{fcutfeoi}
 at $\Delta \delta =$23\arcsec \ might correspond to the \MOLH \
 emission at $\Delta \delta = $ 3\arcsec .   This  lower
 density between ionization and photo--dissociation fronts implies
 a density increase between the atomic and molecular regions by at
 least a factor of 4  since the general bar stratification requires
 an average density of at least $5\times 10^4$  \percc \ in order
 to account for the observed offsets of ionization front and
 molecular lines 
 (see also Wyrowski et al. 1997).  

 Thus, one possible interpretation
 (see also Simon et al. ~\cite{SSSW97})
 of the \MOLH \ data
 is that there is a sharp density gradient
 perpendicular to the line of sight such that the molecular gas has
 much higher density than the partially ionized atomic medium
 adjacent to it. 
 This has the attractive feature that it helps  explain one
 of the puzzles concerning the bar which is the discrepancy 
 (more than an order of magnitude) between
 the hydrogen column density derived by Parmar et al.
 (\cite{PLA91})
 and that inferred by Hogerheijde et al. 
 (\cite{HJD95}) from their \CEIO \
 data.  
 The Parmar et al. data refer essentially to the main H$_2$ bar
 whereas Hogerheijde et al. preferentially sample the fully
 molecular gas to the SE.
 A density gradient may also cause the offset  between
 the ionization front and the main molecular hydrogen peak to
 be larger than that estimated using a homogeneous model whereas 
 molecular hydrogen and carbon radio recombination lines would become
 closer to one another.

 This last aspect
 is particularly interesting in view of the
 coincidence found by Wyrowski et
 al. (1997) between the bar seen in C91$\alpha $ \ emission 
 and the main \MOLH \ bar.    The difficulty in explaining
 this result stems from the fact that one expects the \MOLH \
 emission to come from gas with temperature above 2000 K while the
 carbon line is thought to be
 formed at temperatures which are considerably
 lower. In fact, there is a firm upper limit of 1600 K on the temperature
 of the gas responsible for the C91$\alpha $ \ emission (based on the
 line width).
 The proposed density gradient discussed above may  cause the
 offset between the photodissociation front (i.e \MOLH \
 emission)  and   carbon line emission to diminish. 

A different interpretation of our observations 
is in principle possible. Our main molecular hydrogen peak
{\it C} at $\Delta \delta  =-24\arcsec$ could be produced
by a low velocity shock  preceding the ionization front. 
This would explain the lack of ionization front indicators
coincident with molecular hydrogen. However, the kinematics
of the emission seen in C91$\alpha $ \ by Wyrowski et al.   
are difficult to  explain in this scenario . One needs a
shock velocity of at least 3 \kms \ to excite \MOLH \
(see  Tielens et al. ~\cite{Tea93})  and then 
the observed C91$\alpha $ \ line widths become
difficult to understand.

 It is intriguing that our
\MOLH \  observations do not show any evidence of high density
gas. This is in contrast with the fact that
 the molecular line  data
 (e.g. Tauber et al. 1995, Simon et al.~\cite{SSSW97}, van der Werf et al. 1996) 
 give evidence for a considerable fraction of the gas
 being in clumps with density 
 $\gg 10^5$ \percc . Such clumps can be expected to affect our
 results because most of the lines observed by us are sensitive
 to high emission measure and high density. High density PDRs are
 expected to be hotter and hence considerably more intense in
 \MOLH\ v=2-1 and 1-0 emission than lower-density PDRs
 (cf. Fig.~\ref{f4w}).  
 Nevertheless, our data show no evidence for gas with density
 above $10^5$ \percc \ even towards 
  regions
where Simon et al. (see also van der Werf et al.)
estimate molecular hydrogen densities of order $2\, 10^5$ \percc .
We see no reason on the other hand why clumps should dissipate on a
short timescale when traversing the \MOLH \ dissociation front and
suggest therefore that clumps, while possibly present,
are a secondary phenomenon.

More important in our opinion is the
density gradient mentioned above.  It is worth noting that in the
scenario which we are advocating, the thermal pressure 
may be constant (at a value of the order of
$10^8 $ cm$^{-3}$K) along a line perpendicular
to the bar and we conclude that isobaric models of the Bar are worth
examining.  This incidentally would suggest relatively low      
values for the magnetic pressure and hence magnetic field 
(below 0.5 mG).

\section{Conclusions}
\label{sconcl}
 This study has presented NIR slit spectra of the Orion
 Bar region. 
 Our main result is based on the molecular hydrogen line
 intensities and is that the densities derived from these tracers
 are of order $3-6\times 10^4$ \percc  \ and thus consistent 
 with estimates of the mean density derived from the
 observed stratification of the bar (e.g. Tielens et al.~\cite{Tea93},
 Wyrowski et al. 1997).
 Comparison with the longer wavelength \MOLH \ data
 of Parmar et al.(1991) implies  a tilt for the bar
 relative to the line of sight of $\sim$10 degrees.
 Our data suggest also that the bar may be split into two portions
 along the line of sight  which are separated by 0.2-0.3 parsec.
 It  seems plausible  that the density is 
 lower in the ``atomic'' region (perhaps $2\times 10^4$ \percc )
 than in the molecular gas (of order $10^5$ \percc ).  We
 conclude that models with constant thermal pressure should be
 examined in future studies of the bar region. 
 
 We have  also derived densities for the ionized gas
 in the vicinity of the ionization front using [FeII] line 
 ratios and find values of order $10^4$ \percc . 
 This together with the molecular hydrogen data has convinced
 us that high density neutral clumps play a rather minor role in 
 determining the observed characteristics of the bar.

  A by-product of these observations was that we were able to use
  the observed OI line at 1.317\mic \ as an estimator for the
  ultraviolet radiation field incident upon the bar.
  We  estimate the  normalised FUV intensity G$_{0}$ on the
  bar to be $0.6-3.0\times 10^4$ using this tracer. 

   Finally, we have used the 1.701\mic \ line of He to examine the
   degree of coincidence of helium and hydrogen Str\"{o}mgren
   spheres.  To within our errors, we find that He$^{+}$ 
   and H$^{+}$ coexist and hence that He abundance estimates using
   these tracers should be reliable. 

\begin{acknowledgements}
We are indebted to  D.P. Smits, who provided us 
with the results of his He level
population calculations and J.H. Black for making available to us
his  H$_2$ transition probabilities. Paul van der Werf  allowed
us to use his H$_{2}$ image of the Bar and Alan Moorwood gave
us his IRSPEC data. 
Special thanks are due to Tino Oliva, for his help and comments
on this projects.
This work was partially supported by ASI grant 94-RS-152 and GNA
grant 96/00317 to the Osservatorio di Arcetri.
A.M. acknowledges partial support through GO grant G005.44800 from
Space Telescope Science Institute, which is operated by the Association
of Universities for Research in Astronomy, Inc., under NASA contract
NAS 5-26555.

\end{acknowledgements}

\begin{appendix}
\section{The effects of inclination on the \MOLH \ lines}

\begin{figure}
{\psfig{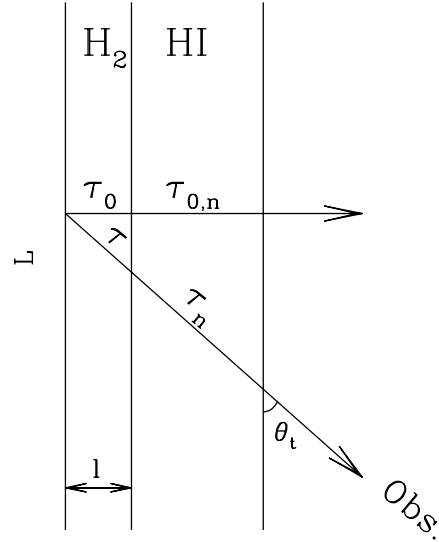}}
 \caption
{ Sketch of the  geometry for molecular hydrogen emission
discussed in the Appendix. The symbols used in Eq.(A1) are
defined. The observer sees the PDR at an angle $\theta _{t}$.
The layer of \MOLH \ emission has continuum optical depth
$\tau _{0}$ and the foreground atomic layer optical depth
$\tau _{0,n}$.}
\label{sketch}
\end{figure}

\begin{figure}
{\psfig{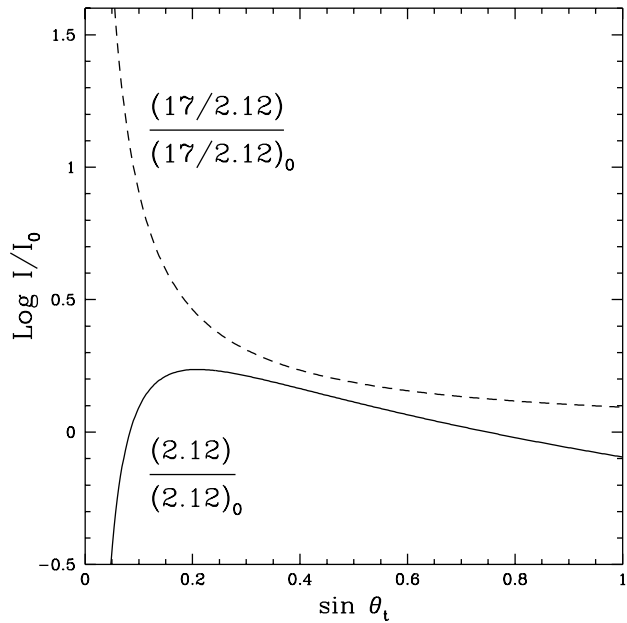}}
 \caption
 {The solid line shows, 
as a function of
the inclination angle $\theta_t$, the intensity of the 1-0 S(1) line 
normalized to the value in a face-on PDR where dust extinction is neglected.
The dashed line shows the ratio
of the two lines 1-0 S(1) and v=0, J=3-1 at 17 \mic, normalized in an
analogous faction.}
\label{figI}
\end{figure}

The  intensity of a given line  as a function of the angle $\theta_t$
can be approximately estimated
from the expression (cf. Fig.~\ref{sketch}):
 \begin{equation}
I=I_0/\tau_0 (1-e^{-\tau})  e^{-\tau_n}
  \end{equation}
where $I_0$ is the face-on intensity ($\theta _{t}=90$)
ignoring extinction due to dust,
$\tau$ the dust optical depth at the 
line frequency across the layer of the PDR which contributes to
the H$_2$ emission,  $\tau_n$ the optical depth of the HI region
along the line of sight. $\tau_0$ and $\tau_{0,n}$ 
are the optical depths
in the \MOLH \  and HI regions, respectively,
in the direction perpendicular to the
PDR. To first approximation, 
$\tau\sim \tau_0/\sin\theta_t$, $\tau_n\sim\tau_{0,n}/\sin\theta_t$.
For  $\theta_t\simless l/L$,
the line of sight does not intercept the neutral region of the PDR
($\tau_n\sim$0) and the line intensity tends to 
$I/I_0\sim 1/\tau_0$. 
When $\tau_0=\tau_{0,n}=0$, then $I/I_0=1/\sin\theta_t$.

Figure ~\ref{figI} shows the ratio $I/I_0$ for the 1-0 S(1) line (solid curve), 
which has been derived using
$\tau_0\sim 1\times A_\lambda/A_v$, $\tau_{0,n}\sim 1\times A_\lambda/A_v$,
$A_\lambda/A_v$=0.145. Here we assume, based upon models, 1 magnitude of visual
extinction in the H$_2$ emitting layer and 1 magnitude of extinction in the
foreground HI layer. For the 1-0 S(1) line,
$I/I_0\sim$ 1 with an accuracy better
than 30\% for $\theta_t\simgreat 8\deg$ . 
The figure shows also the variation with $\theta_t$ of the ratio
of the 1-0 S(1) line at 2.12 \mic\ to the v=0,J=3-1 17\mic\ line (for
which $\tau_0\sim\tau_{0,n}\sim 0$), with respect to the same
ratio for a face-on PDR. 
\end{appendix}



\begin{thebibliography}{}
\bibitem[1991]{Bea91} Baldwin J.A., Ferland, G.J., Martin, P.G.
    et al. 1991, ApJ 374, 580 
\bibitem[1994]{BPO94} Bautista M.A., Pradhan A.K., Osterbrock D.E. 1994,
ApJ 432, L135
\bibitem[1995]{BPD95} Bautista M.A., Pogge R.W., DePoy D.L. 1995,
ApJ 452, 685 
\bibitem[1996]{BD96} Bertoldi F., Draine B. 1996, ApJ 458, 222 
\bibitem[19??]{Cea91} Cardelli J.A. , Clayton G.C., Mathis J.S. 1989,
ApJ 345,245
\bibitem[1994]{DP94} DePoy D.L.,  Pogge R.W.  1994, ApJ 433, 725
\bibitem[1996]{DB96} Draine B.T., Bertoldi F.  1996, ApJ 468, 269 
\bibitem[1995]{DWR95} Dyson J.E., Williams R.J.R., Redman M.P., 1995,
MNRAS 277,700 
\bibitem[1993]{Fea93}
  Felli M., Churchwell E., Wilson T.L., Taylor G.B. 1993,
   A\&AS 98,137 
\bibitem[1993]{FUea93} Fuente A., Martin--Pintado J., Cernicharo J.,
Bachiller R., 1993, A\&A  276,473
\bibitem[1992]{G92} Genzel R. 1992 , in ``The galactic Interstellar Medium",
   p275, (Burton W.B., Elmegreen B.G., Genzel R.), Springer  
\bibitem[1989]{GS89} Genzel R., Stutzki J. 1989, ARA\&A 27,41
\bibitem[1982]{Gea82} Glassgold A.E., Huggins P.J., Schucking E.L.
(Editors) 1982 {\it Symposium  on the Orion Nebula to honor Henry Draper},
 Annals of the New York Academy of Sciences, Vol. 395 
\bibitem[1995]{HJD95} Hogerheijde M.R., Jansen D., van Dishoeck E.F. 1995,
A\&A 294, 792
 \bibitem[1991]{HTT91} Hollenbach D., Takahashi T., Tielens A. 1991, ApJ
                 377, 192
 \bibitem[1995]{HN95} Hollenbach D., Natta A. 1995, ApJ 455,133 
\bibitem[1997]{HT97} Hollenbach D., Tielens A. 1997, ARA\&A (in press) 
\bibitem[1996]{Hea96} Hunt L.K., Lisi F., Testi L.,  et al., 1996, A\&AS 115, 181
\bibitem[1997]{Hea97} Hunt L.K., Migliorini S., Testi L., et al., 1997, AJ (submitted).
\bibitem[1994]{HTBMM94} Hunt L.K., Testi L., Borelli S., Maiolino R., Moriondo G., 1994, Technical Report 4/94, Arcetri Astrophysical Observatory
\bibitem[1995a]{Jea95a} Jansen D.J., van Dishoeck E.F., Black J.H., Spaans M., Sosin C.
1995a, A\&A 302, 223
\bibitem[1995b]{Jea95b} Jansen D.J., Spaans M., Hogerheijde M., van Dishoeck E.F. 
1995b, A\&A 303, 541 
\bibitem[1996]{Lea96} Lisi F., Hunt L.K., Baffa C., et al., 1996, PASP 108, 364
\bibitem[1996]{LCGH96} Lizano S., Canto J., Garay G., Hollenbach D. 1996,
 ApJ 468,739  
 \bibitem[1996]{LJ96} Luhman M.L., Jaffe D.T., 1996, ApJ 463,191
 \bibitem[1995]{M95} Mathis J.S. 1995, {\it Rev Mex AA (Serie de
 Conferencias)} 3, 207 
 \bibitem[1980]{M80} Mezger P.G. 1980 pp81-97 in {\it Radio Recombination
 Lines}, P.A.Shaver (ed), publ. D.Reidel , Vol. 80, Astrophys. and Space
 Science Library 
\bibitem[1991]{Mea91} Moorwood A.F.M., Monet A., Gredel R., 1991, The Messenger 63, 77
 \bibitem[1990]{OMD90} Oliva E., Moorwood A.F.M., Danziger I.J. 1990 A\&A 240,453
\bibitem[1992]{OO92} Oliva E., Origlia L., 1992, A\&A 254, 466
 \bibitem[1990]{OSV90} Osterbrock D.E., Shaw R.A., Veilleux S. 1990, ApJ
  352, 561 
 \bibitem[1992]{OTV92} Osterbrock D.E., Tran H.D., Veilleux S. 1992,
  ApJ 389, 305 
  \bibitem[1979]{PWH79} Pankonin V., Walmsley C.M., Harwit M. 1979,
   A\&A 75,34 
   \bibitem[1991]{PLA91} Parmar P.S., Lacey J.H., Achtermann J.M.
   1991, ApJ 372, L25  
  \bibitem[1982]{P82} Peimbert M. 1982 in {\it Symposium on the Orion 
  Nebula}, edited Glassgold A.E. et al., Annals of New York
  Academy of Sciences Vol 395, 24 
  \bibitem[1993]{P93} Peimbert M. 1993 , {\it Rev. Mexicana de Astron. y Astrof.}
   27,9 
   \bibitem[1995]{P95} Peimbert M. 1995, p163 in {\it The Analysis of Emission
   Lines}, edited R.E.Williams and M.Livio, Space Tel. Science Inst.
   Series, 8 
  \bibitem[1977]{PTP77} Peimbert M., Torres--Peimbert S. 1977,
    MNRAS 179, 217
  \bibitem[1992]{PTR92} Peimbert M., Torres--Peimbert S., Ruis M.T. 1992,
  {\it RevMexAA} 24,155 
   \bibitem[1992]{POA92} Pogge R.W., Owen J.M., Attwood B. 1992, ApJ 399, 147 
   \bibitem[1993]{PZ93} Pradhan A.K., Zhang H.L. 1993, ApJ 409, L77
   \bibitem[1973]{RB73} Robbins, R.R., Bernat, A.P. 1973, Mem.Soc.R.Sci.
      Liege 6th ser., 5, 263
   \bibitem[1996]{Ro96} Rodriguez M. 1996 A\&A 313,L5 
   \bibitem[1991]{Rea91} Rubin R.H., Simpson J.P., Haas M.R., Erickson E.F. 
   1991,  PASP 103, 834 
   \bibitem[1993]{RDW93} Rubin R.H., Dufour R.J., Walter D.K., 1993, ApJ 413,242
   \bibitem[1997]{SSSW97} Simon R., Stutzki J., Sternberg A.,
   Winnewisser G. 1997,  A\&A (in press) 
  \bibitem[1986]{Sea86} Simpson J.P., Rubin R.H., Erickson E.F., Haas M.R.
  1986 , ApJ 311, 895 
 \bibitem[1996]{S96} Smits D.P. 1996, MNRAS 278, 683
 \bibitem[1978]{S78} Spitzer L. 1978, {\it Physical Processes in
 the Interstellar Medium}, publ. J.Wiley \& \ Sons
 \bibitem[1989]{SD89} Sternberg A., Dalgarno A. 1989, ApJ 338,197 
 \bibitem[1995]{SD95} Sternberg A., Dalgarno A. 1995 , ApJS 99, 565
 \bibitem[1995]{SH95} Storey P.J., Hummer D. 1995, MNRAS 272,41
 \bibitem[1997]{SH97} St\"{o}rzer H., Hollenbach D. 1997, ApJ in press.
 \bibitem[1994]{Tea94} Tauber J.A., Tielens A.G.G.M., Meixner M., Goldsmith P.F.
 1994, ApJ 422,136 
\bibitem[1995]{tauber95} Tauber J.A., Lis D.C., Keene J., Schilke P.,  
  B\"uttgenbach, T.H. 1995, A\&A 297, 567
 \bibitem[1985]{TH85} Tielens A., Hollenbach D.  1985, ApJ
                   291, 722      
 \bibitem[1993]{Tea93} Tielens, A.A.G.M., Meixner, M.M., van der Werf, P.P.,
 Bregman, J., Tauber, J.A., Stutzki,J., Rank, D. 1993,  Science
 262,86 
 \bibitem[1977]{Tea77} Turner J., Kirby--Docken K., Dalgarno A.  
  1977 ApJS 35, 281 
 \bibitem[1996]{VdW96} van der Werf P., Stutzki J., Sternberg A., Krabbe A.
1996, A\&A 313, 633
\bibitem[1997a]{Vea97} Vanzi L., Sozzi M., Marcucci G., et al.  1997a A\&AS
   124, 573 
\bibitem[1997b]{VGCT97} Vanzi L., Gennari S., Ciofini M., Testi L., 1997b, Experimental Astronomy, submitted
\bibitem[1995]{VMG95}Vanzi L., Marconi A., Gennari S., 1995, in ``New Developments in Array Technology and Applications'', eds. A.G. Davis Philip et al., p. 231
\bibitem[1997]{W97} Walmsley M. 1997, in  Proceedings if the 1$^{st}$ Arecibo
Workshop on ``Molecular Spectroscopy with the upgraded Arecibo
telescope in the 1-10GHz range'',
editors L.Olmi, W.Baan. , in press
\bibitem[1987]{WJ87} Wilson T.L., J\"{a}ger B. 1987, A\&A 184,291
\bibitem[1990]{WF90} Wilson T.L., Filges L. 1990 in {\it Radio recombination
lines ; 25 years of investigation}, ed. M.A.Gordon, R.L.Sorochenko.
\bibitem[1997]{WFCRR97} Wilson T.L., Filges L., Codella C., Reich W.,
Reich P. 1997, A\&A (in press)
\bibitem[1997]{WSHW97} Wyrowski F., Schilke P., Hofner P., Walmsley M. 1997,
 ApJ 487, L171
\bibitem[1990]{YZ90} Yusuf--Zadeh F. 1990 ApJ 361, L19 

\end{thebibliography}
\end{document}